\documentclass[pdflatex,sn-mathphys-num]{sn-jnl}

\usepackage{graphicx}
\usepackage{multirow}
\usepackage{amsmath,amssymb,amsfonts}
\usepackage{amsthm}
\usepackage{mathrsfs}
\usepackage[title]{appendix}
\usepackage{xcolor}
\usepackage{textcomp}
\usepackage{manyfoot}
\usepackage{booktabs}
\usepackage{algorithm}
\usepackage{algorithmicx}
\usepackage{algpseudocode}
\usepackage{listings}
\usepackage{stfloats}
\usepackage{subfig}
\usepackage{bm}
\usepackage{makecell}
\usepackage{titlesec}

\begin{document}

\title[Article Title]{Near Real-time Full-wave Inverse Design of Electromagnetic Devices}

\author[1]{\fnm{Jui-Hung} \sur{Sun}}\email{juihungs@usc.edu}
\author[1]{\fnm{Mohamed} \sur{Elsawaf}}\email{elsawaf@usc.edu}
\author[1]{\fnm{Yifei} \sur{Zheng}}\email{yzheng@usc.edu}
\author[1]{\fnm{Ho-Chun} \sur{Lin}}\email{hochunli@usc.edu}
\author[1]{\fnm{Chia Wei} \sur{Hsu}}\email{cwhsu@usc.edu}
\author*[1,2]{\fnm{Constantine} \sur{Sideris}}\email{sideris@stanford.edu}
\affil[1]{\small
  {\centering
   \orgdiv{Ming Hsieh Department of Electrical and Computer Engineering,}\par}
  \orgname{University of Southern California},
  \orgaddress{\state{California}, \country{USA}}
}
\affil[2]{\small \orgdiv{Department of Electrical Engineering}, \orgname{Stanford University}, \orgaddress{\state{California}, \country{USA}}}

\abstract{
\small Inverse design enables automating the discovery and optimization of devices achieving performance significantly exceeding that of traditional human-engineered designs. However, existing methodologies to inverse-design electromagnetic devices require computationally expensive and time-consuming full-wave electromagnetic simulation at each iteration or generation of large datasets for training neural-network surrogate models. This work introduces the Precomputed Numerical Green Function method, an approach for ultrafast electromagnetic inverse design. The static components of the design are incorporated into a numerical Green function obtained from a single fully-parallelized precomputation step, reducing the cost of evaluating candidate designs during optimization to only being proportional to the size of the region under modification. A low-rank matrix update technique is introduced that further decreases the cost of the method to milliseconds per iteration without any approximations or compromises in accuracy. This method is shown to have linear time complexity, reducing the total runtime for an inverse design by several orders of magnitude compared to using conventional electromagnetics solvers. The design examples considered demonstrate speedups of up to 16,000x, shortening the design process from multiple days to weeks down to minutes. The approach enables practical and ultrafast design of complex structures that are prohibitively time-consuming for prior inverse design methods.
}


\maketitle

\section*{Introduction}
\label{sec:intro}

Electromagnetic devices are an indispensable part of daily life, playing key roles in telecommunications, radar, sensors, biomedical devices, and more. 
The conventional process of electromagnetic design is heavily reliant on human intuition and experience, and the iterative nature of design is time-consuming and resource-intensive.
As such, inverse design techniques \textemdash algorithmic approaches for the discovery and optimization of devices or structures yielding desired functional properties \textemdash have attracted significant focus across many disciplines, including radiofrequency (RF) or millimeter-wave (mmWave)~\cite{johnson_genetic_1997, altshuler_wire-antenna_1997, robinson_particle_2004, stadler_developing_2022, jin_advances_2007, lalau-keraly_adjoint_2013, zhu_phase--pattern_2021,  naseri_generative_2021, hou_customized_2020, mohammadi_estakhri_inverse-designed_2019, karahan_deep-learning_2024, karahan_deep-learning-based_2023, yang_circularly_2024, 
zhu_frequency_2025}
nanophotonics and optics~\cite{molesky_inverse_2018, li_inverse_2022, hughes_adjoint_2018, peurifoy_nanophotonic_2018, roberts_3d-patterned_2023, ma_deep_2021, camacho_single_2021, piggott_inverse_2015, wiecha_deep_2020, 
grbcic_inverse_2025, sun_edge-guided_2025}
and materials and structural engineering~\cite{sanchez-lengeling_inverse_2018, kumar_inverse-designed_2020, ha_rapid_2023, coli_inverse_2022, zunger_inverse_2018}. 
The properties of interest are encoded as objective functions that are extremized via optimization methods. 
The appeal of the inverse design paradigm owes to its capability for broad exploration of design spaces with many degrees of freedom, enabling the synthesis of novel devices achieving performance superior to that of conventional designs.

In gradient-based inverse design approaches, optimization is performed by iteratively following the gradient of the objective function computed over the space of input parameters.
Such methods are liable to converge to local extrema, and many inverse design runs at random starting configurations may be required before satisfactory results are attained. Moreover, a gradient may not be available due to discrete-valued input parameters, such as metal conductivities and substrate dielectric constants; allowing such parameters to vary continuously may result in physically infeasible designs.
As a result, gradient-free optimization approaches, such as genetic algorithms~\cite{johnson_genetic_1997, altshuler_wire-antenna_1997, molesky_inverse_2018} and particle swarm optimization~\cite{robinson_particle_2004, stadler_developing_2022, jin_advances_2007, kennedy_particle_1995}, have been introduced, enabling wider design space coverage. However, a prominent limitation of both gradient-based and gradient-free techniques for electromagnetic design is that full-wave field simulations are required to evaluate the objective function at each optimization iteration. Even with the fastest commercially-available solvers, such as Ansys HFSS or CST Studio Suite, single simulations often take tens of minutes to hours to run even for structures of only moderate complexity. 

As such, objective function evaluation is typically the rate-limiting factor for design throughput, and mitigating this has been the subject of much work. 
For instance, adjoint methods~\cite{lalau-keraly_adjoint_2013, li_inverse_2022, hughes_adjoint_2018} for gradient-based approaches allow the gradient to be computed with only two field simulations per iteration. 
Alternatively, to dispense with simulation entirely during optimization, machine-learning techniques, which construct surrogate models that allow performance to be predicted from the input parameters, have garnered widespread attention within and beyond electromagnetic design~\cite{zhu_phase--pattern_2021, naseri_generative_2021, hou_customized_2020, peurifoy_nanophotonic_2018, ma_deep_2021,  kumar_inverse-designed_2020, ha_rapid_2023, coli_inverse_2022,karahan_deep-learning_2024, karahan_deep-learning-based_2023, yang_circularly_2024, wiecha_deep_2020, zhu_frequency_2025, grbcic_inverse_2025}. 
While neural-network surrogates can greatly reduce optimization time, the overall process is constrained by computationally-expensive training as well as the large number of simulations needed for adequate design space coverage when generating training datasets. Although approaches such as transfer learning have been introduced to enhance training efficiency, the training phase, inclusive of the generation of the large dataset (on the order of 10,000 to $>$ 1 million simulations
\cite{zhu_phase--pattern_2021, hou_customized_2020, peurifoy_nanophotonic_2018, coli_inverse_2022, karahan_deep-learning_2024}), 
may nonetheless require multiple days to weeks 
\cite{naseri_generative_2021, karahan_deep-learning_2024, yang_circularly_2024, wiecha_deep_2020}. 
Furthermore, over the full design space, there is no guarantee of accuracy given valid input parameters~\cite{alizadeh_managing_2020, wiecha_deep_2020}; that is, a design predicted as optimal by the surrogate may yield completely different performance when verified with full-wave simulation or measurement of a fabricated device.

This paper introduces the precomputed numerical Green function (PNGF) method, an approach for inverse design of electromagnetic devices with rapid, direct, approximation-free objective function evaluation at each optimization iteration. 
Structure(s) to be optimized in the design are replaced with equivalent electric current densities, allowing the interaction between the static, unchanging portions of the device and the dynamic optimization region to be represented by a numerical Green's function matrix obtained from a single fully parallelized precomputation step. 
Once found, the numerical Green's function may be used to evaluate the performance of any candidate device within the design environment.

Reusing computations to accelerate solutions to electromagnetics problems is a widely used paradigm. For example, domain decomposition techniques \textemdash speeding up the analysis of a computational domain by solving subdomains and coupling adjacent subdomains together with boundary conditions at the interfaces \textemdash 
are often accelerated by reusing previously computed quantities. 
Applications have included perturbation approaches with surface currents in a method-of-moments (MoM) formulation for analyzing large aperiodic antenna arrays \cite{ludick_efficient_2014}, optimization of photonic crystal structures with a Dirichlet-to-Neumann discretization \cite{verweij_accelerating_2014}, and speeding up the simulation of large metallic platforms with local variations \cite{gao_rapid_2022}, among others. 
While the PNGF method shares broadly similar motivations, it is distinct from domain decomposition approaches. The numerical Green's function reduces the problem size to only the number of unknowns in the optimization region, and the remainder of the domain, including the static portions of the device, is incorporated into the numerical Green's function itself.

While this work demonstrates PNGF on pixelated metallic structures using the direct binary search (DBS) global optimization algorithm,
the method is readily applicable to dielectric optimization problems and can also be used with other optimization approaches. 
Many optimizers, including DBS, make a small modification to the design at each iteration. As such, low-rank update methods may be utilized to accelerate objective function evaluation significantly without trade-offs in accuracy. 
The cost of evaluation is linear with the size of the optimization region, which reduces the total runtime of the full inverse design by multiple orders of magnitude, from several days or weeks (with commercial solvers) to minutes. 
Unlike neural network-based surrogate models, the PNGF method yields a highly-accurate solution, which matches those obtained from the full-wave electromagnetic solver used for precomputation with multiple digits of precision and is correct for every design in the feasible set, without training.

The PNGF method is applied to design three example devices: an ultrawideband 30GHz substrate antenna with 40\% fractional bandwidth, a 6GHz planar switched-beam antenna (SBA) whose beam is switchable over a $70^\circ$ angle, and a broadband short-length transition between a microstrip feedline and a substrate-integrated waveguide (SIW). When PNGF is utilized as the solver for DBS, speedups of up to four orders of magnitude in the optimization time versus DBS using the fastest commercial software (e.g., HFSS and CST) are obtained. The SBA and microstrip-SIW transition are fabricated, and the measured scattering parameters of the devices and the radiation pattern of the SBA agree closely with predicted simulation results.

\section*{Results}
\subsection*{Current equivalence}
\label{sec:currequiv}

An electromagnetic device to be optimized typically encompasses a predefined optimization region, where the goal of design is to find a configuration of structure(s) in the optimization region that yields desired electromagnetic properties. Additional components of the device, such as dielectric structures, ground planes, feedlines, and air gaps, that lie outside the optimization region are constrained to be static and are excluded from being modified.
To model the device, 3D space is discretized into a set of voxels (typically by using a finite-difference or finite-element algorithm) in a simulation environment that encompasses the device and its immediate surroundings, such as an air box, as well as models of excitation source(s), such as lumped ports and incident fields.

Each optimization iteration would be significantly accelerated if candidate designs could be assessed as linear combinations of solutions to simpler designs. However, such solutions do not obey superposition. Although a structure can be viewed as a combination of simpler parts, the sum of the fields scattered by each part alone (under the same excitation) would not match the fields of the full structure due to scattering interactions between substructures.
As such, traditional optimization approaches have required full-wave simulations to reevaluate the entire environment, including the static components, from scratch at each iteration. 

\begin{figure}[!ht]
\centering
\includegraphics[width=\textwidth,trim=10 10 10 10,clip]{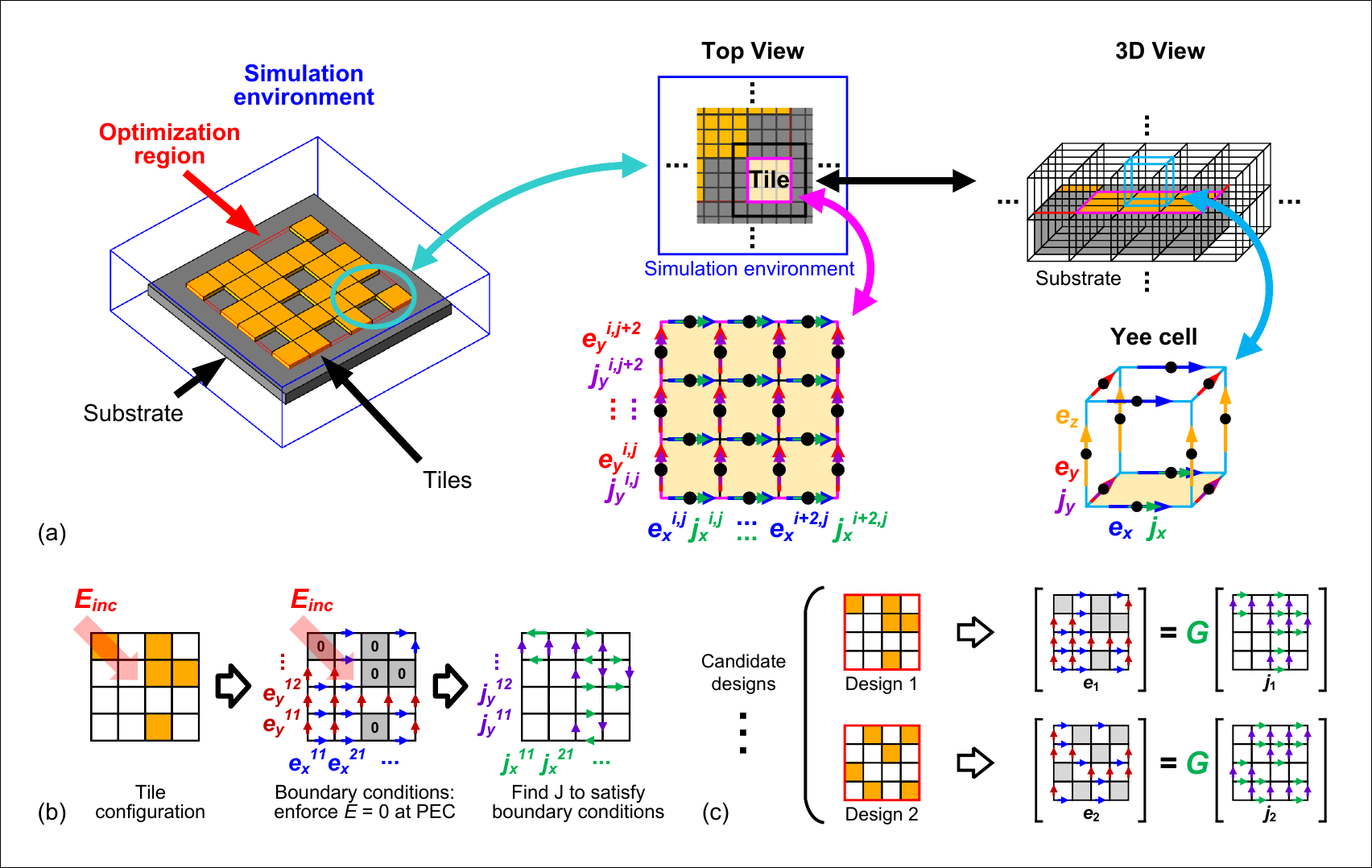}
\caption{\textbf{Current equivalence principle applied to pixelated electromagnetic devices.} 
(a) Example discretization of a representative structure with planar optimization region, where each voxel is a finite-difference Yee cell and each tile comprises the faces of $3 \times 3$ cells; 
(b) Process to replace an arbitrary arrangement of metallic tiles with equivalent current densities that satisfy boundary conditions and produce identical fields;
(c) Concept of a numerical Green's function matrix, allowing equivalent current densities and corresponding fields to be found for arbitrary configurations of the optimization region to satisfy boundary conditions.
For simplicity, tiles in (b) and (c) are shown as comprising one voxel each, but in practice, multiple voxels constitute a tile.
}
\label{fig:doms_curreq}
\end{figure}

To overcome this limitation, we apply the current equivalence theorem to represent any given configuration of the optimization region with polarization densities, which create fields identical to those that would be generated by the original device in response to an excitation. 
Arbitrary current densities and the fields that they produce are linear and obey superposition. This
allows candidate designs to be evaluated by solving a linear system whose number of unknowns is equal to the size of the optimization region only, as opposed to the full simulation environment. The interactions between the optimization region and the static components of the design are encoded into a numerical Green's function, a matrix that maps current densities to electric fields in the optimization region.

We introduce the current equivalence concept using pixelated metallic devices, in which a grid of tiles, each of which may be filled with metal or left open, forms the optimization region. Such devices are readily modeled via finite-difference methods, where each tile comprises a rectangular array of Yee cells, as shown in Fig. \ref{fig:doms_curreq}(a), and the polarization densities and electric fields found with current equivalence are co-located on the edges of Yee cells. The concept of solving for the unknown polarization densities with a candidate design is depicted in Fig. \ref{fig:doms_curreq}(b), and utilizing the numerical Green's function is shown in Fig. \ref{fig:doms_curreq}(c). Although we only consider metal in our optimization region for practical demonstration, this technique can be used with any arbitrary complex permittivity distribution to represent lossy metals and/or dielectrics, as discussed in Supplementary Note \ref{supp:pngf_curreq_ext}.

\subsection*{Numerical Green's functions}
\label{sec:ngf}

A list of symbols used throughout this work is provided in Supplementary Table \ref{tb:supp_listofsymbols}.
We seek to replace any given configuration of the optimization region with an equivalent effective polarization density
$\mathbf{J_p}(\mathbf{r}) = \varepsilon_0 (\varepsilon_r(\mathbf{r}) - 1) \mathbf{E}(\mathbf{r})$, 
where $\varepsilon_0$ is the free-space permittivity and $\varepsilon_r(\mathbf{r})$ is the material permittivity,
such that the fields $\mathbf{E}$ produced by $\mathbf{J_p}$ in an empty optimization region are identical to those with the original metallic tiles in response to an excitation $\mathbf{E_{inc}}$. 
The conductivity $\sigma(\mathbf{r})$ at points $\mathbf{r}$ throughout the optimization region is either zero (free space) or infinity (metal represented by perfect electrical conductor (PEC) material). 
Defining an auxiliary quantity $p(\mathbf{r})$ such that $\sigma(\mathbf{r}) = \frac{p(\mathbf{r})}{1-p(\mathbf{r})}$, it can be shown (see Supplementary Note \ref{supp:ce_deriv}) using the electric field volume integral equation~\cite{markkanen_discretization_2012} that
\begin{equation}
p(\mathbf{r}) \mathbf{E_\text{inc}} = 
(1- p(\mathbf{r})) \mathbf{J_p(\mathbf{r})} + 
p(\mathbf{r}) \int_V \overline{\overline{G_0}}(\mathbf{r}, \mathbf{r'}) \mathbf{J_p}(\mathbf{r'}) \; dV'
,
\label{eq:currequiv_j_cont}
\end{equation}
where $\overline{\overline{G_0}}$ is the dyadic free space Green's function, and regions $\mathbf{r}$ in the domain $V$ with $p(\mathbf{r})=1$
correspond to metal and $p(\mathbf{r})=0$ correspond to free space. The solution $\mathbf{J_p}$ is unique and results in zero tangential electric field wherever there is metal, satisfying the PEC boundary conditions.

Equation (\ref{eq:currequiv_j_cont}) is strictly valid when the design comprises only metallic tiles and vacuum, as it uses the dyadic free-space Green's function.
However, $\overline{\overline{G_0}}$ may be replaced with the Green's function for any particular simulation environment, where additional materials representing arbitrary dielectric or metallic structures (e.g., substrates or feed lines) outside the optimization region are generally present. 
While closed-form analytical Green's functions are usually not available, it is known that a Green's function exists for every linear system, and a discrete numerical Green's function matrix $G$ can be obtained using a full-wave EM solver.

\begin{figure}[!ht]
\centering
\includegraphics[width=\textwidth,trim=10 10 10 10,clip]{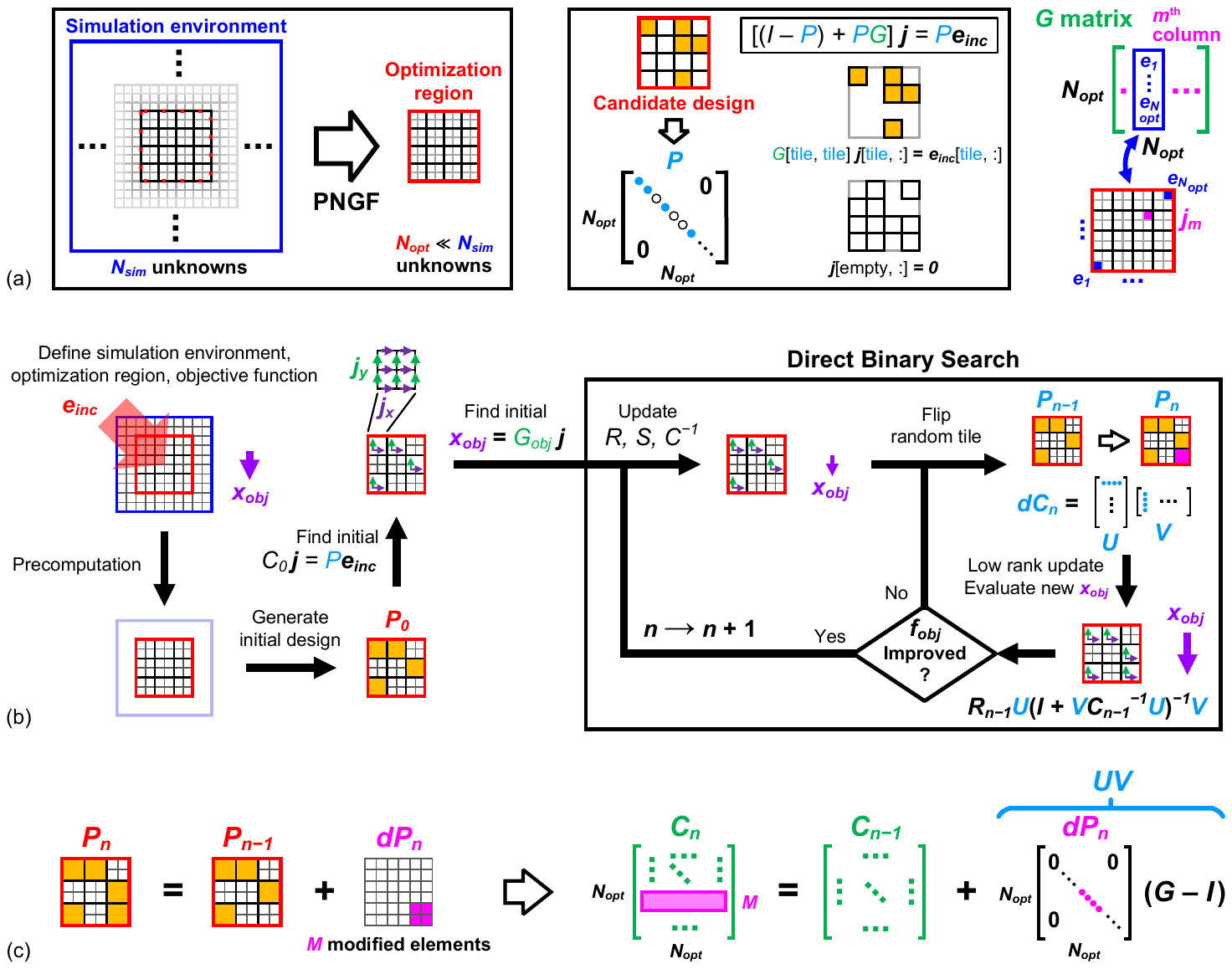}
\caption{\textbf{Precomputed numerical Green function optimization with direct binary search.} 
(a) Numerical Green function matrix $G$ allowing candidate designs $P$ to be evaluated by solving a linear system of only $N_{opt}$ unknowns; 
(b) Process of direct binary search optimization with the PNGF method;
(c) Tile flip yielding a low-rank update to the PNGF system matrix, which is performed with the Woodbury matrix identity in this work. 
For simplicity, tiles are shown as comprising 2x2 voxels each, whereas tiles generally encompass more voxels in practice. The illustration of $G$ at the right of (a) is a simplified representation with one unknown per voxel.
}
\label{fig:sys_dbs_lru}
\end{figure}

To solve for the current density numerically for a given design under consideration, 
by choosing a suitable basis to represent $\mathbf{J_p}$ (e.g., rooftop functions) and testing functions~\cite{harrington_method_1987}, equation (\ref{eq:currequiv_j_cont}) can be discretized into
\begin{equation}
[(I-P) + PG]\mathbf{j} = C\mathbf{j} = P\mathbf{e_\text{inc}}
,
\label{eq:currequiv_j_disc}
\end{equation}
where $\mathbf{j}$ and $\mathbf{e_\text{inc}}$ are the discretized polarization density and incident electric field vectors over the optimization region, and the design is encoded in $P$, a diagonal matrix whose entries indicate metal ($1$) or an empty location $(0)$. 
The vector indices correspond to the degrees of freedom in the optimization region as discretized with the solution method of choice (e.g., spatial points on the Yee grid when a finite-difference method is utilized), with a total of $N_{opt}$ such degrees of freedom. 
The matrix $G$, a discretized form of the Green's function integral operator, needs only to be precomputed once for a given simulation environment, and then any candidate design may be evaluated by solving the linear system of equations (\ref{eq:currequiv_j_disc}) over only the optimization region, as illustrated in Fig. \ref{fig:sys_dbs_lru}(a). 
The number of unknowns $N_{opt}$ is considerably smaller than the number of unknowns $N_{sim}$ comprising the full simulation environment. This involves no approximations and incurs no loss of accuracy compared to a conventional full-wave simulation of the full system. 

\subsection*{Precomputation}
\label{sec:precomp}
 
In general, an electromagnetic field solver finds the inverse of a matrix $A$ that satisfies $A\mathbf{e_{sim}} = \mathbf{j_{sim}}$, where the electric fields $\mathbf{e_{sim}}$ and currents $\mathbf{j_{sim}}$ encompass the entire simulation environment. 
The matrix $A$ corresponds to the discretized Maxwell operator (e.g., the finite-difference frequency-domain matrix) of the simulation environment, comprising the static region and an empty optimization region (Supplementary Note \ref{supp:fdfd_apf}). 
However, to obviate the need to compute the full $A^{-1}$, a tall logical $0-1$ projection matrix $B$ may be defined to map vectors $\mathbf{e_{opt}}$ and $\mathbf{j_{opt}}$ in the optimization region to the corresponding vectors in the full simulation environment; that is, $\mathbf{e_{opt}} = B^T \mathbf{e_{sim}}$ and $\mathbf{j_{sim}} = B \mathbf{j_{opt}}$, where $B^T B = I$. 
The matrix $B^T$ extracts the fields from the appropriate indices in the full simulation result $\mathbf{e_{sim}}$ corresponding to the optimization region, and $B$ maps the currents in the optimization region $\mathbf{j_{opt}}$ into their corresponding locations in the larger $\mathbf{j_{sim}}$ vector, filling the remaining entries with zeroes.
From the definition of $B$, it follows that
\begin{equation}
\mathbf{e_{opt}} 
= B^T A^{-1} \mathbf{j_{sim}} 
= B^T A^{-1} B \mathbf{j_{opt}}
.
\label{eq:precomp_syseq_dom}
\end{equation}
The matrix $B^T A^{-1} B$ corresponds to $G$ in equation (\ref{eq:currequiv_j_disc}), as it maps currents $\mathbf{j_{opt}}$ to fields $\mathbf{e_{opt}}$ in the optimization region, while considering the static parts of the simulation outside of the optimization region via the full system matrix $A$. 

An iterative solver may be used to obtain $G$ column-by-column, where each simulation yields the fields due to a current density at a single discretized spatial location in the optimization region (i.e., each column of $B$). The $N_{opt}$ simulations are completely independent of each other and, as such, may be run in parallel across many nodes. Additionally, time-domain methods such as finite-difference time-domain (FDTD) may be used, where the frequency-domain information in $G$ is obtained from discrete Fourier transforms after each simulation. 
For multi-frequency optimization, this allows columns of multiple $G$ matrices (one at each desired frequency) to be obtained from a single FDTD simulation, provided that the excitation has a sufficiently wide bandwidth to excite all frequencies of interest.

Alternatively, a sparse direct solver may be used with a frequency-domain formulation to obtain $G$ efficiently in a single shot with the recently-introduced augmented partial factorization (APF) technique~\cite{lin_fast_2022}. The full system matrix $A$ is constructed using the finite-difference frequency-domain (FDFD) formulation (see Supplementary Note \ref{supp:fdfd_apf}). Then, an augmented sparse matrix $K$ is set up such that $A$ comprises the upper left block. $K$ can be partially factorized as
\begin{equation}
    K =
    \begin{bmatrix}
        A & B \\
        B^T & 0
    \end{bmatrix}
    =
    \begin{bmatrix}
        K_{L11} & 0 \\
        K_{L21} & I
    \end{bmatrix}
    \begin{bmatrix}
        K_{U11} & K_{L12} \\
        0 & K_{U22}
    \end{bmatrix}
    ,
\end{equation}
where $K_{L11}$ and $K_{U11}$ are the LU-factors and $K_{L21}$ and $K_{L12}$ are additional blocks not used for precomputation. 
The matrix $K_{U22}$, known as the Schur complement~\cite{zhang_schur_2005}, is given by $K_{U22} = -B^T A^{-1} B$. 
The Schur complement has been previously applied to electromagnetics problems to reduce the size of the simulation domain, such as in domain decomposition methods as with \cite{verweij_accelerating_2014, gao_rapid_2022}.
The $G$ matrix is obtained as $-K_{U22}$, avoiding the need to apply the LU factors for $A$ to find $B^T A^{-1} B$.

Field quantities outside the optimization region are often required to evaluate the objective function. A vector $\mathbf{x_{obj}}$ of quantities needed for evaluation may be defined, and a matrix $G_{obj}$ that maps current densities in the optimization region to $\mathbf{x_{obj}}$ may be precomputed and utilized in optimization together with $G$. 
Each of the $N_{obj}$ elements of $\mathbf{x_{obj}}$ is a field value or a linear combination of field values. In practical scenarios, $N_{obj}$ is rarely significantly larger than $1$; for example, to compute a scalar mode amplitude with a discrete mode overlap integral, $\mathbf{x_{obj}}$ would have $N_{obj} = 1$ element that is a linear combination of the fields at the evaluation points~\cite{lumerical_modeoverlap}, and for a near-field to far-field transformation, two linear combinations are needed~\cite{taflove_computational_2005}, giving $N_{obj} = 2$.
A wide logical $0-1$ projection matrix $W^T$ may be defined to obtain $\mathbf{x_{obj}}$ from the full simulation environment solution $\mathbf{e_{sim}} = A^{-1} B \mathbf{j_{opt}}$:
\begin{equation}
\mathbf{x_{obj}} = W^T \left(A^{-1} B \mathbf{j_{opt}} + \mathbf{e_{inc}}\right) = G_\text{obj} \mathbf{j_{opt}} + \mathbf{x_{inc}},
\label{eq:objeval_ft_def}
\end{equation}
where $G_{obj}=W^TA^{-1}B$ and $\mathbf{x_{inc}}=W^T\mathbf{e_{inc}}$. Using the above methods, $G_\text{obj}$ may be precomputed together with $G$ directly without additional computational cost. With an iterative solver, the number of excitations to be solved is still $N_{opt}$, since $G_{obj}$ has $N_{opt}$ columns. If APF is employed, the augmented system becomes
\begin{equation}
    K_{obj} =
    \begin{bmatrix}
        A & \begin{bmatrix} B & 0\end{bmatrix} \\
        \begin{bmatrix} B^T \\ W^T \end{bmatrix} & 0
    \end{bmatrix}
    ,
\end{equation}
and the Schur complement yields the vertical concatenation of $G$ and $G_{obj}$ with a single run of the solver. Since the sparse direct matrix solver requires a square matrix, the $B$ matrix block in the augmented system is padded with columns of zeroes corresponding to the number of rows of $W^T$. 

It should be noted that although we have discussed obtaining the numerical Green's function using FDTD and FDFD methods, any solution method of choice can be used, including finite-element and integral equation methods (e.g., Supplementary Note \ref{supp:alt_discr}).
Once precomputation has been performed for a given simulation environment and discretization scheme, $G$ and $G_{obj}$ may be used for any number of optimization runs with the same environment.

\subsection*{Direct binary search optimization}
\label{sec:optimization}

The Direct Binary Search (DBS) optimization algorithm starts with an initial design configuration $P_0$, which may be randomly-generated or based on a priori design insight. 
The inverse of the initial system matrix, $C_0^{-1} = \left[(I-P_0) + P_0 G\right]^{-1}$, is found and stored, and the objective function is evaluated. 
At each iteration, a randomly chosen tile in the optimization region is flipped from empty to filled or vice versa. The objective function is evaluated, and if 
improvement is obtained,
the flip is retained and optimization proceeds to the next iteration. Otherwise, another random tile is flipped. Should all possible flips be tested without improvement, the optimization has converged. The DBS process utilizing PNGF is illustrated in Fig. \ref{fig:sys_dbs_lru}(b), and a flowchart is shown in Supplementary Fig. \ref{fig:supp_flowchart}.

At the $n$th iteration of optimization,
\begin{equation}
    \mathbf{x_{{obj,} n}} =
    G_\text{obj} \mathbf{j_{{opt},n}}+\mathbf{x_{inc}} = G_\text{obj} C_{n}^{-1} P_n \mathbf{e_{inc}} + \mathbf{x_{inc}}
    \label{eq:xobj_definition}
\end{equation}
must be found to evaluate the objective function. 
Although $C_n^{-1} = [(I - P_n) + P_n G]^{-1}$ could be obtained by solving equation (\ref{eq:currequiv_j_disc}) from scratch,
since DBS flips only a single tile per iteration, low-rank update methods can be used instead to avoid recomputing the inverse directly, as illustrated in Fig. \ref{fig:sys_dbs_lru}(c) and discussed in the following section.

It is important to note that the update methods introduced in the following section are not restricted to DBS. Indeed, many optimization algorithms (e.g., levelset methods, as discussed in Supplementary Note \ref{supp:levelset}) modify a small ($\ll N_{opt}$) number of unknowns in the design matrix $P$ at each iteration of inverse design, and the low-rank update techniques are readily applied to such approaches.

\subsection*{Low-rank update evaluation}
\label{sec:lru}

After a tile flip, the number $M$ of modified elements in the diagonal design matrix $P$ is, in the case where a finite-difference discretization is used, the number of field components comprising a tile on the Yee grid. Let $dP_n = P_n - P_{n-1}$ represent the change to $P$.  A wide logical $0-1$ projection matrix $Q$ may be constructed such that $dP_n = Q^T H_{P,n} Q$, where $H_{P,n}$ is a diagonal $M$-by-$M$ matrix whose entries are the nonzero elements of $dP_n$. Let $U = Q^T H_{P,n}$ and $V = Q(G-I)$. Then, the update $dC_n = C_n - C_{n-1}$ may be expressed as
\begin{equation}
dC_n = dP_n(G-I) = \left[Q^T H_{P,n}\right]\left[Q (G - I)\right] = UV
.
\label{eq:lru_dcn}
\end{equation}

Since $C_n^{-1}$ is the inverse of a low-rank (rank-$M$) update to $C_{n-1}$, the Woodbury matrix identity~\cite{bayin_mathematical_2006} may be employed to obtain $C_n^{-1}$ through a rank-$M$ update to the inverse of $C_{n-1}$. 
In computational electromagnetics, the Woodbury identity has seen use for efficient low-rank updates to a system matrix, particularly with method of moments-based formulations. Prior work has employed the identity for applications including
efficient computation of the gradient of an objective function via perturbations to the impedance matrix for optimization of metasurfaces \cite{budhu_fast_2023}, 
photonic crystal design through low-rank updates to a Wannier basis analysis of the device \cite{jiao_demonstration_2005}, 
and inverse design with a method of moments perturbation approach based on removing basis functions \cite{capek_shape_2019}.

Using the Woodbury identity,
\begin{equation}
C_n^{-1} = 
\left(C_{n-1} + UV\right)^{-1} =
C_{n-1}^{-1} - C_{n-1}^{-1} U \left( I + V C_{n-1}^{-1} U\right)^{-1} V C_{n-1}^{-1}
.
\label{eq:lru_woodbury}
\end{equation}
However, for many tile flips, the objective function will be worse than that of the previous iteration, and $C_n^{-1}$ would be discarded once found. 
Further performance may be obtained by instead finding $\mathbf{x_{obj, n}}$ directly using $C_{n-1}^{-1}$. Substituting equation (\ref{eq:lru_woodbury}) into $\mathbf{x_{{obj,} n}} = G_\text{obj} C_{n}^{-1} P_n \mathbf{e_{inc}}+ \mathbf{x_{inc}}$ yields 
\begin{equation}
\mathbf{x_{obj, n}} = G_\text{obj}
\left[C_{n-1}^{-1} - C_{n-1}^{-1} U \left( I + V C_{n-1}^{-1} U\right)^{-1} V C_{n-1}^{-1}\right]
P_{n} \mathbf{e_{inc}}+\mathbf{x_{inc}}
.
\label{eq:lru_eobj_uv}
\end{equation}
Let 
\begin{align}
    R_{n-1} &= G_\text{obj} C_{n-1}^{-1} 
    , \label{eq:lru_r}\\
    S_{n-1} &= C_{n-1}^{-1} P_{n-1}\mathbf{e_{inc}} 
    , \label{eq:lru_s} \\
    \mathbf{x_{obj, n-1}} &= G_\text{obj} C_{n-1}^{-1} P_{n-1} \mathbf{e_{inc}}  + \mathbf{x_{inc}} \label{eq:lru_xinc}
    .
\end{align}
Equation (\ref{eq:lru_eobj_uv}) becomes
\begin{align}
\begin{split}
\mathbf{x_{obj, n}} = & \left(R_{n-1}dP_n \mathbf{e_{inc}}+\mathbf{x_{obj,n-1}}\right) \\
&-
R_{n-1} U 
\left( I + V C_{n-1}^{-1} U \right)^{-1} V \left(C_{n-1}^{-1} dP_n \mathbf{e_{inc}}+S_{n-1}\right)
.
\label{eq:lru_eobj_rs}
\end{split}
\end{align}
Since $R_{n-1}$, $S_{n-1}$, and $\mathbf{x_{obj, n-1}}$ do not depend on the current candidate design $P_{n}$, they may be computed once at the start of a new iteration (after each successful tile flip) and used to rapidly evaluate $\mathbf{x_{obj, n}}$ for new flips until the objective function is improved. Once this occurs, the tile flip is retained, $C_{n}^{-1}$ of the following iteration is obtained with equation (\ref{eq:lru_woodbury}), and $R_{n-1}$ and $S_{n-1}$ are updated. It can be shown (Supplementary Note \ref{supp:pngfcost}) that computing $\mathbf{x_{obj, n}}$ for a tile flip and updating $C^{-1}$ for a successful tile flip cost $O(N_{opt})$ and $O(N_{opt}^2)$ operations, respectively, owing to the sparsity of $dP$ and $U$.

\subsection*{Computational efficiency}
\label{sec:compeff}

The cost of objective function evaluation due to flipping a tile is completely independent of the size ($N_{sim}$) or complexity of the overall simulation domain and grows only linearly with respect to the number of unknowns ($N_{opt}$) inside the optimization region. 
Thus, for an optimization region of fixed size, the evaluation cost remains unchanged regardless of the surrounding static environment outside the optimization region.
If a tile flip improves the objective function, updating the system matrix using equation (\ref{eq:lru_woodbury}) requires $\mathcal{O}(N_{opt}^2)$ operations, which is significantly fewer than the $\mathcal{O}(N_{opt}^3)$ needed to invert $C_n$ directly and also substantially smaller than the $\mathcal{O}(N_{sim}^3)$ operations required to invert the original full electromagnetic system. 

\begin{figure}[!ht]
\centering
\includegraphics[width=\textwidth,trim=10 3 10 10,clip]{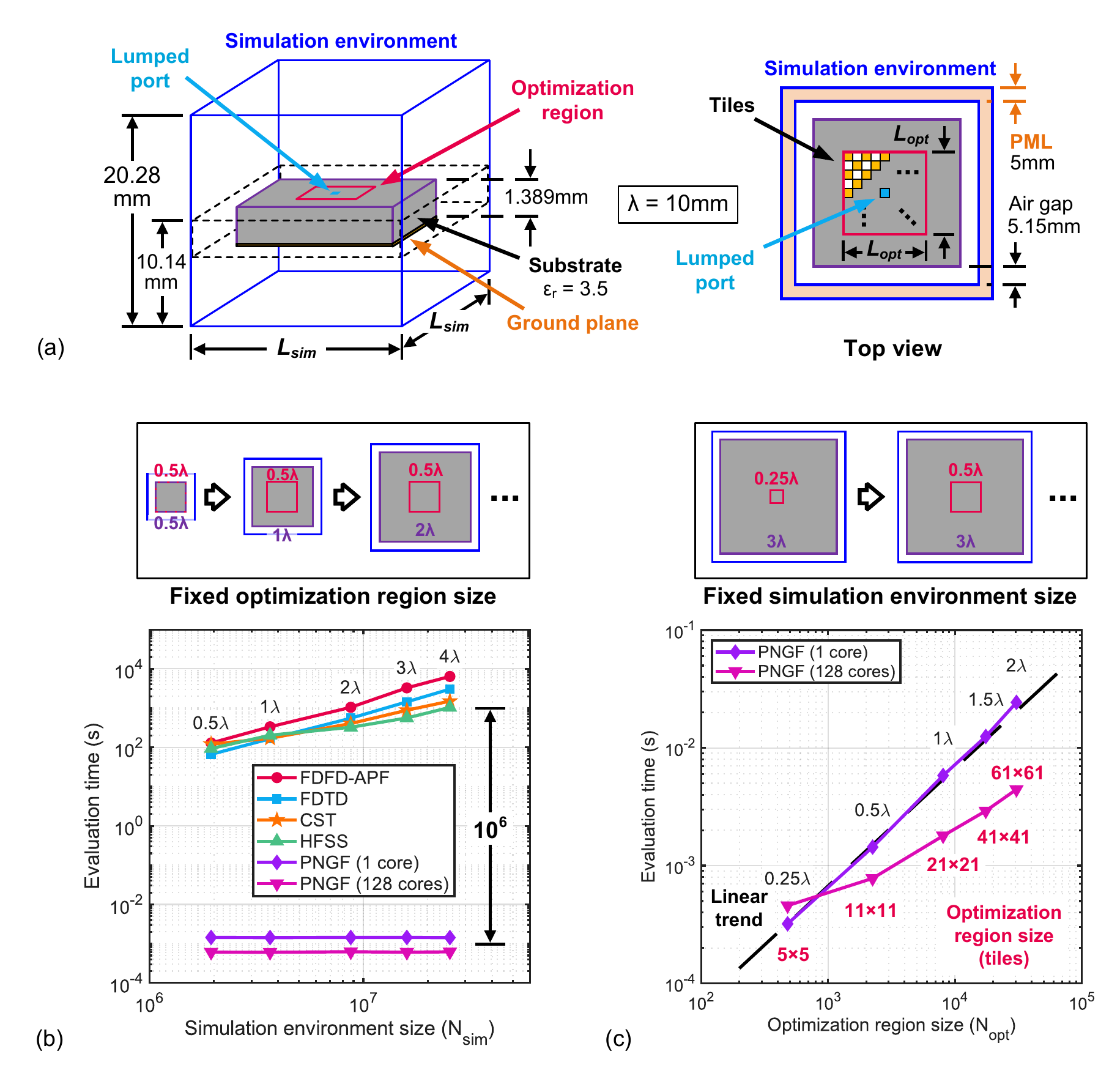}
\caption{\textbf{Runtime performance of the precomputed numerical Green function method.} 
(a) Simulation environment for benchmarking objective function evaluation using PNGF, where the optimization region is populated with tiles ($3 \times 3$ voxels each) in a checkerboard pattern;
(b) Performance of PNGF compared to full-wave electromagnetic solvers versus simulation environment size with a fixed optimization region ($0.5\lambda \times 0.5\lambda$), where PNGF is constant-time;
(c) Performance of PNGF versus optimization region size for a fixed simulation environment  ($3\lambda \times 3\lambda$), demonstrating linear runtime with respect to the optimization region size. Note the 128-core times appear sublinear since the larger size problems better utilize all of the available CPU cores.
}
\label{fig:objeval}
\end{figure}

Two examples to illustrate this performance are presented in Fig. \ref{fig:objeval}.
First, a planar optimization region of fixed size is considered on the surface of a dielectric substrate, and 
the evaluation time
versus the simulation domain size ($N_{sim}$) is plotted for PNGF and compared with full-wave electromagnetic solvers as the substrate and simulation dimensions are increased. 
The second case keeps the simulation ($N_{sim}$) and substrate size fixed and plots the evaluation time versus the optimization region size ($N_{opt})$. 
The simulation environment is a 3D region with a finite dielectric substrate ($\lambda_0 = 10$mm, $\varepsilon_r=3.5$, thickness: $1.389$mm) defined such that its sides are spaced at a fixed distance ($\lambda_0/2$) from Perfectly Matched Layer (PML) absorbing boundary layers~\cite{taflove_computational_2005}. 
The optimization domain is a square region centered on the substrate, and a ground plane covers the substrate bottom. Tiles ($0.5 \times 0.5$mm, $3 \times 3$ Yee cells) populate the optimization region in a checkerboard pattern, a 2D lumped port ($3 \times 2$ Yee cells) is defined in the center, and the objective function is the reflection coefficient.
In comparison to FDFD (using APF as a solver), a custom FDTD solver, HFSS \cite{ref_hfss}, and CST \cite{ref_cst}, PNGF achieves ultra-fast ($<100$ms for all problem sizes tested) performance, faster than all other approaches by multiple orders of magnitude (10,900x{\textendash}1,680,000x for the simulation sizes in Fig. \ref{fig:objeval}(b)).
Furthermore, PNGF provides an approximation-free solution, in common with the full-wave solvers considered, that is accurate for every possible optimization region configuration.
For any given design, the PNGF results match with multiple digits of precision with those obtained by the solver used for precomputation (see Supplementary Note \ref{supp:numerical_accuracy}).

If multiple frequencies are of interest in optimization, a $G$ matrix, with corresponding $G_{obj}$ and $C_0^{-1}$, may be precomputed for each frequency. The low-rank update evaluation procedure may then be applied to each system. Since each system is independent, finding each $\mathbf{x_{obj}}$ after attempted tile flips and updating each $C$ matrix after successful flips may be performed in parallel without any communication overhead.

It should be noted that, as long as $G$ represents the simulation environment for the discretization scheme utilized for the optimization region and $G_{obj}$ obtains the optimization region quantities of interest for evaluating the objective function, there are no limitations on the use of equation (\ref{eq:lru_eobj_rs}) for efficient updates to $x_{obj,n}$. In particular, there is no upper limit on $M$. However, when $M \sim N_{opt}$, the evaluation of equation (\ref{eq:lru_eobj_rs}) may be slower than directly inverting $C$ and using that to obtain $x_{obj,n}$ with equation (\ref{eq:xobj_definition}).

\begin{table}[h]
\caption{Comparison of the runtime of direct binary search inverse design using the precomputed numerical Green function method as a solver versus Ansys HFSS and CST Studio Suite.}
\label{tb:performance}
\begin{tabular}{*5c}
    \toprule
    & & 
        $\begin{matrix}\textbf{Substrate} \\ \textbf{antenna}\end{matrix}$ & 
        $\begin{matrix}\textbf{Switched-beam} \\ \textbf{antenna}\end{matrix}$ & 
        $\begin{matrix}\textbf{Substrate-integrated} \\ \textbf{waveguide}\end{matrix}$ \\
    \toprule
    \multicolumn{2}{c}{\textbf{Number of tile flips}} &
        1051 &
        1939 &
        617 \\
    \midrule
    \multicolumn{2}{c}{$\begin{matrix}\textbf{Number of} \\ \textbf{optimization frequencies}\end{matrix}$} &
        5 &
        3 &
        5 \\
    \midrule
    \textbf{HFSS} & 
        Optimization\footnotemark[1] & 
            $\begin{matrix} 186\text{h} \\ (10.6\text{min} \times 1051) \end{matrix}$ &
            $\begin{matrix} 472\text{h} \\ (14.6\text{min} \times 1939) \end{matrix}$ &
            $\begin{matrix} 134\text{h} \\ (13.1\text{min} \times 617) \end{matrix}$ \\
    \midrule
    \textbf{CST} & 
        Optimization\footnotemark[1] & 
            $\begin{matrix} 106\text{h} \\ (6.05\text{min} \times 1051) \end{matrix}$ &
            $\begin{matrix} 104\text{h} \\ (3.23\text{min} \times 1939) \end{matrix}$ &
            $\begin{matrix} 134\text{h} \\ (13.0\text{min} \times 617) \end{matrix}$ \\
    \midrule
    \multirow{11}{*}{\textbf{PNGF}} &
        $\begin{matrix}\text{Precomputation}\\\text{(FDTD)}\end{matrix}$ &
            $\begin{matrix} 29.5\text{min} \\ (20.6\text{s} \times 86) \end{matrix}$ &
            $\begin{matrix} 89.6\text{min} \\ (33.2\text{s} \times 162) \end{matrix}$ &
            $\begin{matrix} 180\text{min} \\ (41.8\text{s} \times 129) \end{matrix}$ \\ \cmidrule(l){2-5}  
        & $\begin{matrix} \text{Precomputation}\\\text{(APF)}\end{matrix}$ &
            7.82min &
            7.66min &
            83.5min \\ \cmidrule(l){2-5} 
        & $\begin{matrix}\text{Inverting initial}\\\text{system matrix}\footnotemark[2]\end{matrix}$ &
            $\begin{matrix} 63.1\text{s} \\ (12.6\text{s} \times 5) \end{matrix}$ &
            $\begin{matrix} 33.1\text{s} \\ (11.0\text{s} \times 3) \end{matrix}$ &
            $\begin{matrix} 19.1\text{min} \\ (229\text{s} \times 5) \end{matrix}$ \\ \cmidrule(l){2-5} 
        & Optimization\footnotemark[2] &
            $\begin{matrix} 88.6\text{s} \\ (17.7\text{s} \times 5) \end{matrix}$ &
            $\begin{matrix} 22.7\text{s} \\ (7.56\text{s} \times 3) \end{matrix}$ &
            $\begin{matrix} 137\text{s} \\ (27.4\text{s} \times 5) \end{matrix}$ \\ \cmidrule(l){2-5} 
        & $\begin{matrix}\text{Average}\\\text{iteration time}\end{matrix}$ &
            84ms &
            12ms &
            222ms \\ \cmidrule(l){2-5} 
        & Total (using APF) &
            10.3min &
            8.59min &
            105min \\
    \midrule
    \multicolumn{2}{c}{\textbf{Speedup}\footnotemark[3]} &
        4,310x &
        16,600x &
        3,510x \\
    \botrule
\end{tabular}
\footnotetext[1]{Total optimization times for these commercial solvers are estimates obtained by extrapolating the runtime of a simulation of the final optimized designs using the number of attempted tile flips in DBS design with PNGF.}
\footnotetext[2]{For evaluating runtime performance, inverting the initial system matrix, evaluating the objective function, and performing the low-rank update when the objective function improves are performed sequentially for each of the optimization frequencies. However, since each system at each frequency is independent, each of these steps may be parallelized readily.}
\footnotetext[3]{The speedup figure compares only the optimization time, since precomputation for the PNGF method needs only to be performed once for a given simulation environment, and the precomputed Green function matrices may be reused for any subsequent optimization runs. The smaller of the HFSS and CST times is used for each case.} 
\end{table}

\subsection*{Design studies}
\label{sec:designstudies}

For each design study, PNGF is performed with two precomputation approaches: iterative, employing a custom GPU-accelerated FDTD solver, and direct, utilizing APF. 
The frequencies at which optimization is performed and the objective functions for each case are detailed in Supplementary Note \ref{supp:objfuncs}.
Simulations to verify the final designs are performed with HFSS and a custom FDTD solver. A comparison of the runtime performance of DBS optimization for each design study utilizing PNGF compared to full-wave solvers is shown in Table \ref{tb:performance}. The total times for PNGF represent the duration needed to design each device from scratch, with no requisite prior training or pre-existing libraries of simulated designs.

\begin{figure}[!ht]
\centering
\includegraphics[width=\textwidth,trim=10 10 10 10,clip]{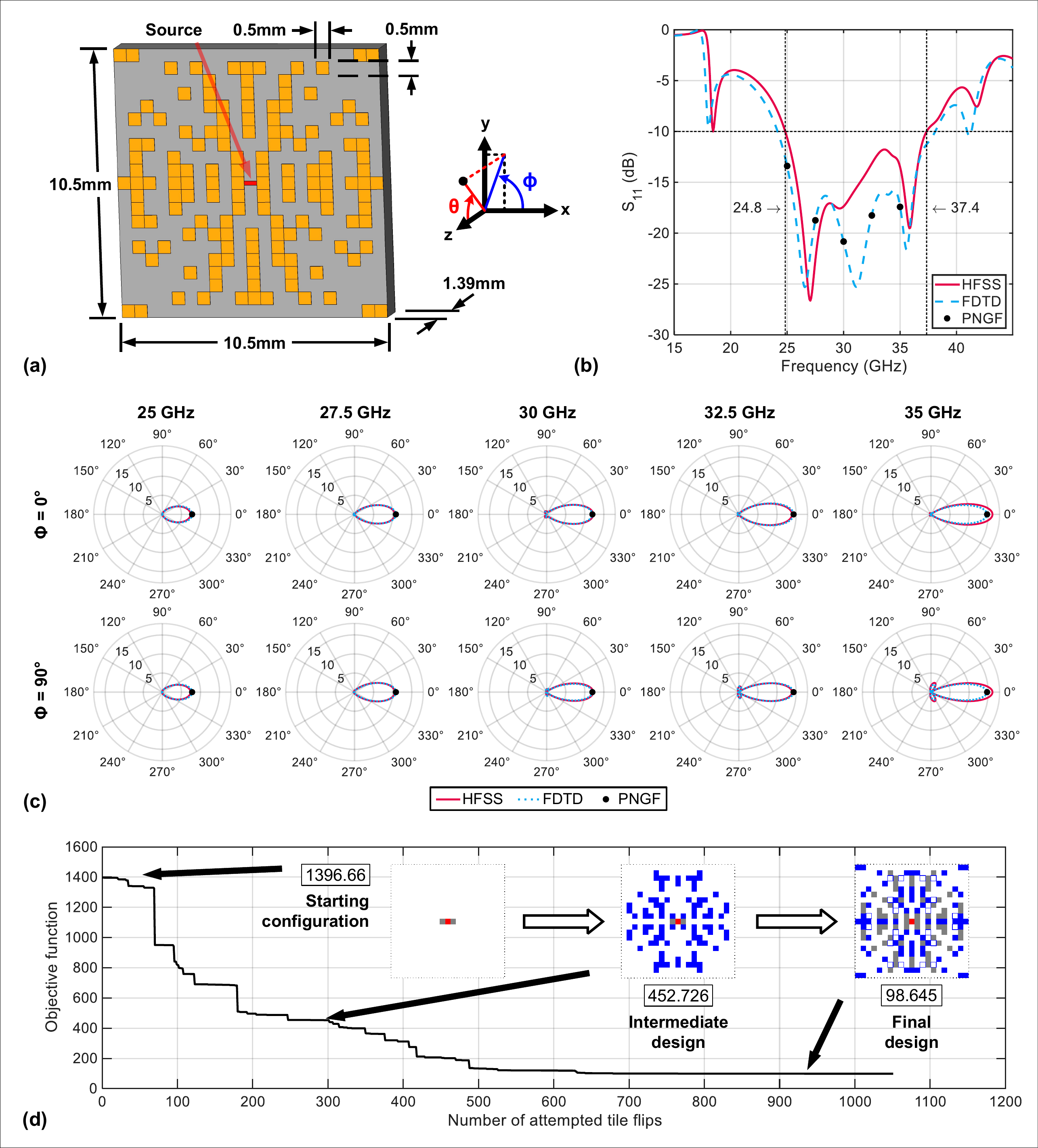}
\caption{\textbf{Broadband 30GHz substrate antenna.} (a) Antenna design with indicated dimensions; (b) Simulated $S_{11}$ with HFSS and FDTD with values from the final iteration of PNGF; (c) Simulated radiation patterns at frequencies spanning the bandwidth in linear scale relative to an ideal isotropic radiator; (d) Evolution of objective function during inverse design.
}
\label{fig:subant}
\end{figure}

Progress in wireless systems, such as ultrawideband technologies, sub-terahertz/terahertz communications, and Internet of Things, has placed ever-increasing demands on antenna capabilities, performance, and size~\cite{kimionis_printed_2021, wu_universal_2023, zhao_2d_2023, ahmad_compact_2022, zheng_ultra-fast_2023, islam_highly_2022, liu_compact_2013, liu_miniaturized_2017, kibaroglu_64-element_2018, dadgarpour_one-_2016, karmokar_fixed-frequency_2016, yang_compact_2019, mackertich-sengerdy_tailored_2023}. 
Planar antennas have been the subject of particular interest~\cite{kimionis_printed_2021, wu_universal_2023, zhao_2d_2023, ahmad_compact_2022, zheng_ultra-fast_2023, liu_miniaturized_2017, karmokar_fixed-frequency_2016} 
owing to their ease of fabrication and integration where space is limited.
However, traditional topologies, such as patch antennas, are often narrowband or have strongly frequency-dependent radiation patterns.
We design a broadband 30GHz center-fed substrate antenna with a wide fractional bandwidth and highly uniform pattern. 
The design is shown in Fig. \ref{fig:subant} with simulated reflection coefficient and radiation patterns. The performance obtained at the final iteration of PNGF is also plotted, showing agreement with finite-difference simulation (see Supplementary Note \ref{supp:numerical_accuracy})
The design exhibits a 10dB return loss bandwidth of approximately 13GHz, corresponding to a 40\% fractional bandwidth. 
The radiation pattern remains largely unchanged over the entire frequency range, where the gain in the broadside direction is greater than 8.7 dBi with a peak of 12.1dBi at 35GHz.

\begin{figure}[!ht]
\centering
\includegraphics[width=\textwidth,trim=10 3 10 10,clip]{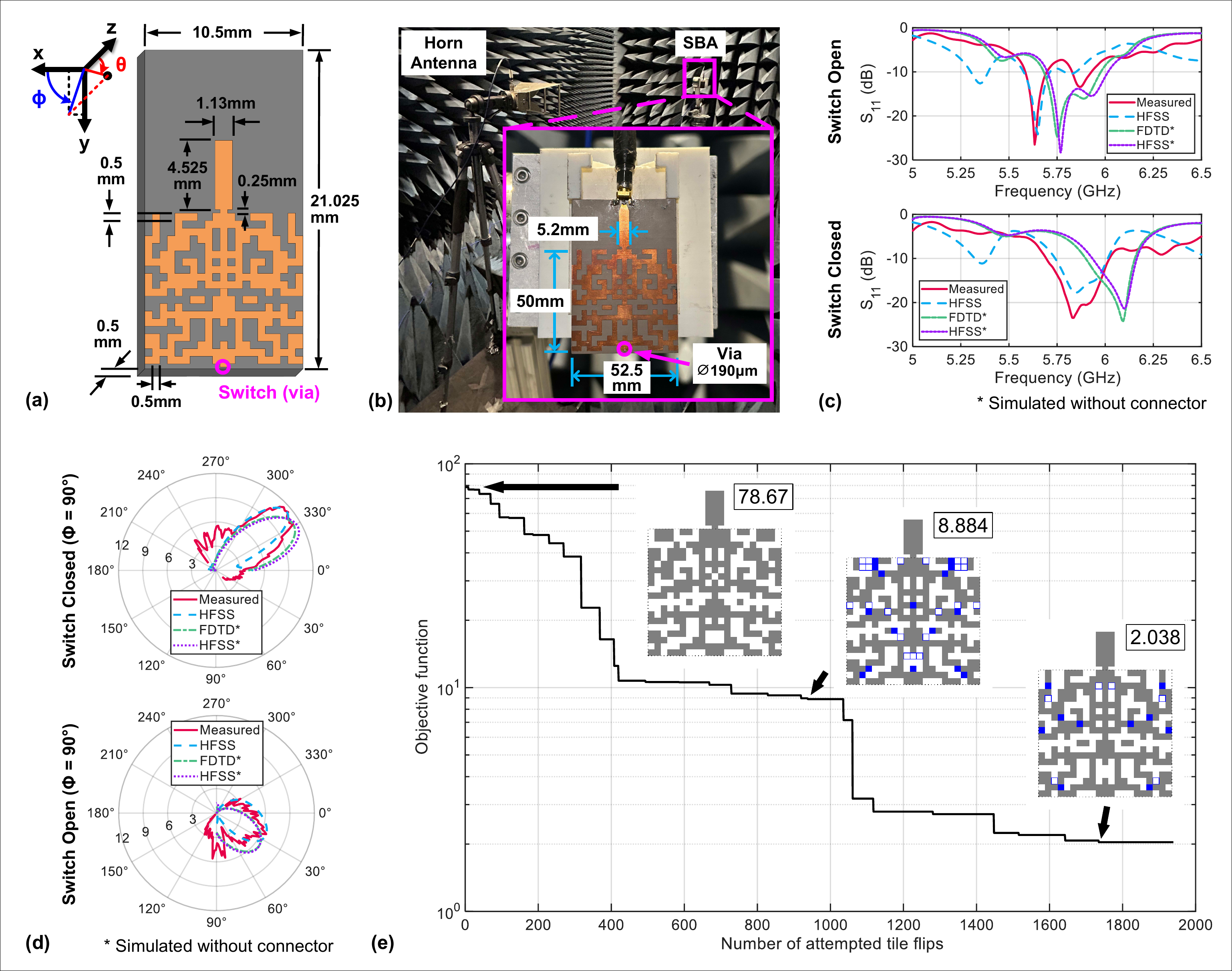}
\caption{\textbf{Reconfigurable switched-beam antenna for 5G cellular applications.} (a) 30GHz design with indicated dimensions; (b) Fabricated scaled 6GHz antenna with feed and 2.92mm connector on measurement setup; (c) Simulated and measured $S_{11}$ of 6GHz design with the switch open and closed; (d) Simulated and measured radiation patterns of 6GHz design with the switch open and closed at $\phi = 90^\circ$ (yz plane), in linear scale relative to an ideal isotropic radiator; (e) Evolution of objective function during inverse design. 
The measured pattern is normalized to the maximum gain of the simulation results.
A slight deviation in the simulated $S_{11}$ and patterns with HFSS and FDTD arises because the connector is not modeled in the FDTD simulation; HFSS simulation without the connector demonstrates excellent agreement with FDTD.}
\label{fig:sba}
\end{figure}

Advancements in cellular networks have placed ever-growing requirements on antennas for transmitting and receiving multidirectional, ultrawideband signals ~\cite{zhao_2d_2023, islam_highly_2022, ahmad_compact_2022, liu_compact_2013, liu_miniaturized_2017, kibaroglu_64-element_2018, dadgarpour_one-_2016}. 
The multiple-input multiple-output functionality of current cellular technology is typically realized using phased arrays or multiple antenna elements
\cite{ahmad_compact_2022, liu_compact_2013, liu_miniaturized_2017, kibaroglu_64-element_2018}, whose large electrical size restricts miniaturization. 
As such, reconfigurable antennas, whose properties may be altered dynamically with inputs (e.g., switches), have garnered substantial attention
\cite{dadgarpour_one-_2016, karmokar_fixed-frequency_2016, yang_compact_2019, mackertich-sengerdy_tailored_2023}. 
We design a 30GHz switched-beam antenna (SBA) for 5G applications, with a switch for selecting between two target beam directions ($\theta = 45^\circ, \phi = 90^\circ$ with switch open; $\theta = 45^\circ, \phi = 270^\circ$ with switch closed). 
Due to practical equipment limitations, we scaled the inverse-designed antenna up in all dimensions by a factor of 5x before fabrication to shift the center frequency to 6GHz and facilitate measurement. 
The fabricated SBA is shown in Fig. \ref{fig:sba} with the simulated and measured reflection coefficient and radiation patterns versus $\theta$ for $\phi = 90^\circ$. 
The simulations are performed with the fabricated upscaled design, and the measurements agree closely, with a 10dB return loss bandwidth of 0.3GHz (switch closed) and simulated peak gains of 8.2dBi (switch open) and 10.6dBi (closed). The measured angle of beam switching when viewed in the yz plane is approximately $70^\circ$.

\begin{figure}[!ht]
\centering
\includegraphics[width=\textwidth,trim=10 3 10 10,clip]{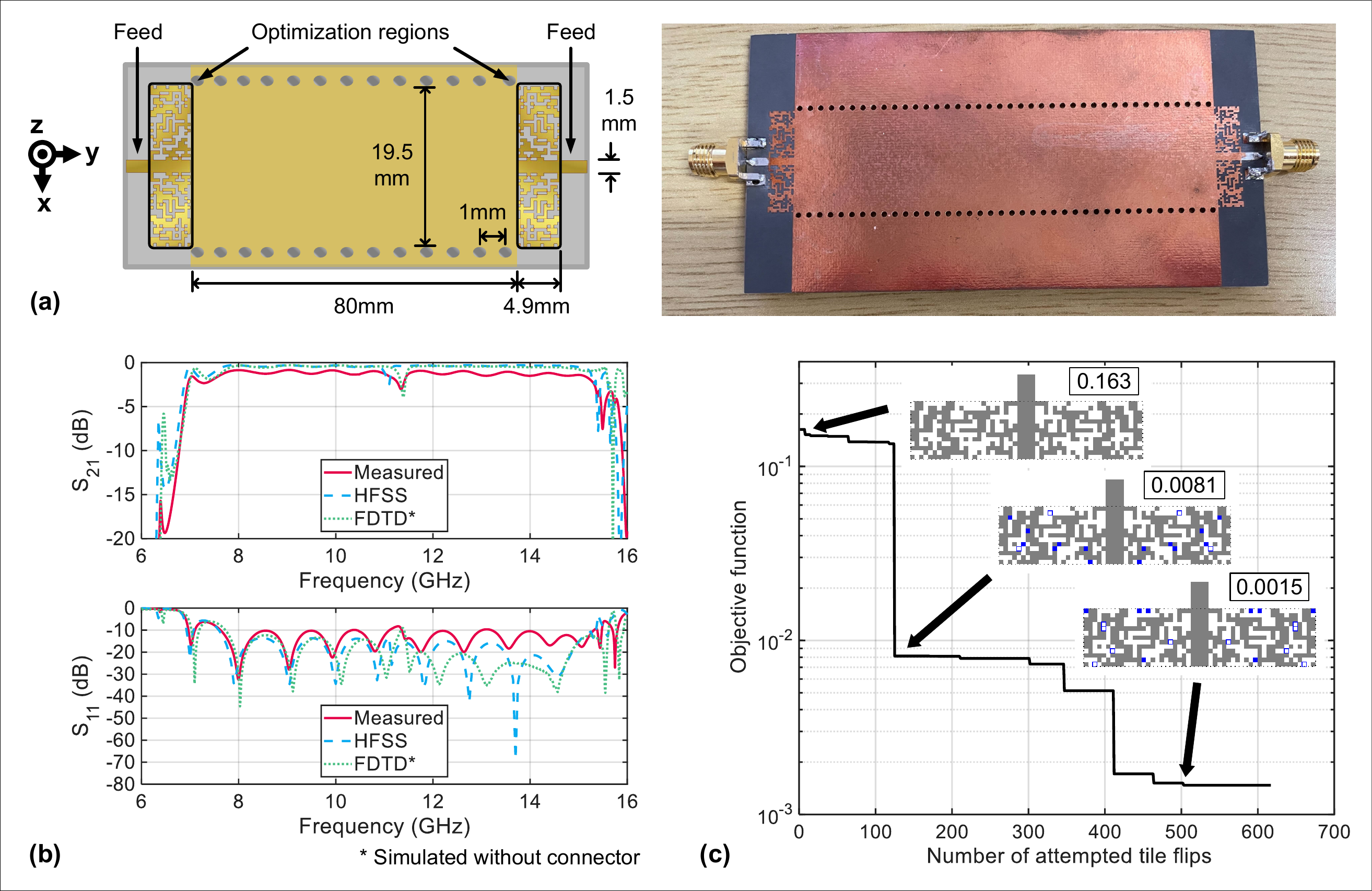}
\caption{\textbf{Broadband 8\textendash15GHz transition from microstrip transmission line to substrate-integrated waveguide.} (a) Waveguide section with transitions, microstrip feeds, and 2.92mm connectors on both ends; (b) Simulated and measured $S_{11}$ and $S_{21}$ of the designed structure; (c) Evolution of objective function during inverse design.
}
\label{fig:siw}
\end{figure}

Substrate-integrated waveguides (SIWs) comprise a planar substrate enclosed by metal cladding and side walls formed by vias. Owing to their compatibility with printed circuit board fabrication processes, SIWs have attracted considerable interest~\cite{uchimura_development_1998,deslandes_accurate_2006,deslandes_integrated_2001,lee_v-band_2006,elsawaf_concurrent_2023}. The fundamental mode is typically excited with a tapered transition from a microstrip feed to the SIW. For compactness, it is desirable to decrease the length of the transition, but this often yields decreased performance. We design a taperless transition from a 50$\Omega$-impedance transmission line to a broadband SIW. The length of the optimization region is more than 4x shorter than the length of a linear taper required to achieve comparable bandwidth, as in~\cite{elsawaf_concurrent_2023}. Optimization is performed to minimize the insertion loss over the bandwidth of interest. 
The fabricated waveguide section with transitions and the simulated and measured $S_{11}$ and $S_{21}$ are shown in Fig. \ref{fig:siw}, demonstrating a wide 10dB return-loss bandwidth of approximately 7.7GHz.
As with the preceding SBA design, minor differences between the simulated (with feed connector) and the measured results are visible. The discrepancy likely arises from uncertainties and variations in fabrication and material properties, such as permittivity, substrate losses, substrate thickness, and surface roughness of the metallization, as well as imperfections associated with the measurement setup and instrumentation.

\section*{Discussion}
\label{sec:discussion}

A new approach for the inverse design of electromagnetic structures has been presented. 
By encapsulating the static, unchanging components of the design into a numerical Green function matrix, the method allows any candidate design to be evaluated by solving a linear system with only as many unknowns as the size of the optimization region in the design environment. 
When utilized with the direct binary search optimization algorithm or other optimization strategies that also perform sparse updates to the optimization domain at each iteration, a low-rank update technique can be employed to further accelerate objective function evaluation at each iteration, achieving linear time complexity with respect to the size of the optimization domain. 

Runtime improvements up to six orders of magnitude are demonstrated without compromising on accuracy when compared to state-of-the-art commercial solvers, such as Ansys HFSS and CST Studio Suite, and three high-performance design examples, relevant to contemporary RF/wireless technologies, are demonstrated and experimentally verified. 
Using the PNGF method, the full inverse design process, inclusive of the precomputation phase, was on the order of single hours or less for all design examples considered. It took approximately 10 minutes in total to inverse design structures with optimization regions of approximately $1 \lambda$ by $1 \lambda$,
whereas approaches using conventional solvers may take multiple days to weeks. Considering the optimization time alone, all of the examples took less than 140 seconds to design.
The PNGF method achieves speeds competitive with AI-based surrogate models, but does not require any training and is guaranteed to produce the correct solution for any candidate design input.

Future work includes considering multilayer problems, including on-chip filters, matching networks, and other passives, extending the approach to dielectric problems for nanophotonic applications, investigating other optimization algorithms, such as levelset methods and particle swarm optimization, and leveraging alternative solvers to precompute the PNGF matrix, such as integral equation methods and finite element methods. 
As discussed in Supplementary Notes, the PNGF method can be readily generalized for optimization regions containing dielectric materials (\ref{supp:pngf_curreq_ext}) and employed with levelset methods (\ref{supp:levelset}) and discretization schemes other than finite-difference methods such as method of moments (\ref{supp:alt_discr}). Additionally, with dielectric problems, the gradient of the objective function may be efficiently obtained for gradient-based optimization (\ref{supp:gradient}). 
Although the PNGF method was developed and applied in the context of electromagnetics in this work, we expect that the approach can readily be adapted to any scenario that can be modeled by a linear system, such as heat transfer and acoustic wave propagation. 
With broad applicability and validated performance, the PNGF method has transformative potential for the inverse design of electromagnetics structures.

\section*{Methods}
\label{sec:methods}

\subsection*{Finite-difference discretization}
\label{meth:fd_discr}

For each test case in Fig. \ref{fig:objeval} and each design study, the same Yee lattice is used for both FDTD and APF precomputations as well as the optimization with PNGF. The optimization regions are rectangular areas made up of faces of adjoining Yee cells. Equivalent polarization density components, as discussed in the Results section (Precomputation), in x and y are defined on these faces. For example, with tiles comprising $3 \times 3$ Yee cell faces as illustrated in Fig. \ref{fig:doms_curreq}(b), each tile comprises 12 $j_x$ and 12 $j_y$ components, and the number of modified elements $M$ in a tile flip is $24$. The current components are co-located with the electric field x and y components on the edges of the Yee cells. Such flat optimization regions are appropriate for modeling the copper-clad utilized in these examples, whose thickness is much smaller than the wavelength. For other applications, however, the optimization region may encompass multiple layers of Yee cells, within which z-components of the current density would also be present. The simulation environment for each design is truncated using Perfectly Matched Layer absorbing boundaries.

\subsection*{Computational resources}
\label{meth:comp_resources}

A custom solver that constructs the FDFD system and performs APF is used to generate the $G$ and $G_{obj}$ matrices. 
The MUMPS package \cite{amestoy_fully_2001} is used to carry out partial factorization and compute the Schur complement.
For large simulation environments, it may be infeasible to use a direct solver owing to the required amount of random access memory, and iterative solvers must be employed, incurring the cost of running $N_{opt}$ full-wave simulations. However, this is not the case for the three design studies considered in this work, and APF achieves up to 12x decreases in execution time compared to the GPU-accelerated FDTD for precomputation.

For each design study, APF precomputations are run on three 128-core AMD EPYC 7763 nodes, where each precomputation uses 64 cores on one node. These are run in parallel, one for each frequency of optimization in each design.
For GPU-accelerated FDTD precomputations, 24 FDTD simulations are run in parallel using the EPYC 7763 nodes with one Nvidia A100-SXM4-80GB GPU per simulation. For timing comparison, HFSS and CST are run with 128 cores on one node, and the PNGF implementation utilizes the BLAS and LAPACK linear algebra packages with the same number of CPU cores. As optimization is performed at multiple frequencies for each case, evaluation with PNGF is performed sequentially for each frequency during each iteration.
For HFSS, the HFSS solution type (finite-element method-based) is used with a multi-frequency solution at each optimization frequency with the direct solver with first-order basis functions and adaptive meshing. For CST, the time domain solver (finite integration technique-based) with a hexahedral mesh, which is similar to FDTD, is used without adaptive meshing.

For the objective function evaluation benchmarking of Fig. \ref{fig:objeval}, HFSS and CST are utilized as solvers using the same resources as above, respectively. The FDTD solver is a custom multithreaded implementation using OpenMP and 128 cores on a single node. FDFD is performed with APF as a solver, using 128 cores on a single node. PNGF is used for objective function evaluation for each case with a single core and also with 128 cores on a single node.

\subsection*{Optimization parameters}
\label{meth:op_params}

The substrate antenna design utilizes a 1.39mm-thick substrate ($\varepsilon_{r}=3.5$ representing Rogers RO3035) cladded with 13.9{\textmu}m-thick copper. The bottom copper layer is fully filled as a metal ground reflector, and the optimization region is defined on the top layer. The optimization region comprises a 21x21 grid of tiles, where each tile is 3$\times$3 Yee cells of 0.5 $\times$ 0.5mm each. This results in 4032 $e_x$/$j_x$ and 4032 $e_y$/$j_y$ components. 
In view of maximizing the gain in the broadside direction, x and y symmetry are enforced; as such, only 2048 simulations are required when precomputations are performed with FDTD, and 4 tiles are flipped at a time during optimization. 
A 50$\Omega$ x-directed lumped port in the center is used as the excitation source for field simulations. 

The 30GHz SBA design utilizes a 0.508mm-thick Rogers TC350 ($\varepsilon_r = 3.5$) cladded with 18{\textmu}m-thick copper. The bottom layer comprises the ground plane whereas the top design domain is approximately one wavelength square with 21$\times$20 tiles each comprising 3$\times$3 Yee cells of 0.5 $\times$ 0.5mm each, giving 3843 $e_x$/$j_x$ and 3840 $e_y$/$j_y$ components. With axial symmetry, the number of simulations necessary for precomputations with FDTD is 3872, and 2 tiles are flipped per optimization iteration. The switch is modeled as an ideal metallic via connecting the top and the bottom. The antenna is edge-fed with a microstrip feed; for field simulations, a z-directed 50$\Omega$ lumped port is attached between the feed and the bottom ground. The objective function is set to minimize the reflection coefficient and increase the directivity in the directions of interest for each configuration of the switch. 

The microstrip-SIW transition design utilizes a 0.508mm-thick Rogers RT/duroid 5880 substrate ($\varepsilon_{r} = 2.2$) with 35{\textmu}m-thick copper clad. The bottom layer is completely filled with copper as a ground plane. Each of the two optimization regions, which are constrained to be identical, comprises 52$\times$13 tiles. Since the fundamental mode and the structure are longitudinally symmetric, the optimization region can be reduced in half, to 26 $\times$ 13. Each tile comprises 3x3 Yee cells of 0.125mm $\times$ 0.125mm each, corresponding to a total of 6240 $e_x$/$j_x$ and 6123 $e_y$/$j_y$  components in the optimization region, and 6201 simulations are required for FDTD precomputations. 
In field simulations, a z-directed 50$\Omega$ lumped port is attached to each microstrip feed end, connecting the ground plane (bottom layer) to the microstrip (top). To design a broadband device, the objective function is set to minimize the insertion loss at five different frequencies. 

\subsection*{Device fabrication}
\label{meth:device_fab}

The scaled 6GHz SBA is fabricated from a 2.5mm-thick Rogers TC350 substrate with $\varepsilon=3.5$ cladded with 1oz copper. To implement the reconfiguration switch of the fabricated SBA, two antennas are fabricated which differ only in whether the switch via is present (switch closed) or absent (open). A 2.92mm end-launch RF connector (Withwave SM03FS017) is soldered to pads at the end of the microstrip feed of each antenna to provide excitation.

The SIW section is fabricated from 0.508mm thick RT/duroid 5880 laminate cladded with 1oz copper. Two 2.92mm end-launch RF connectors (Withwave SM03FS007) are soldered at both microstrip feed ends for the $S_{11}$ and $S_{21}$ measurements.

\subsection*{Measurement system}
\label{meth:measurement}

To measure the reflection coefficient and radiation pattern of the SBA, a measurement setup is established in an anechoic chamber with a vector network analyzer (VNA) (Keysight N5247). For pattern measurements, the SBA is affixed to a two-axis rotary positioner (Diamond Engineering DCP252) driven by stepper motors, and an excitation horn antenna (Com-Power AH-118) is positioned facing the SBA. The pattern is obtained by recording transmission coefficient data with the VNA connected to both antennas, while the elevation and azimuth are swept.

The $S_{11}$ and $S_{21}$ measurements of the microstrip-to-SIW transition are obtained with a VNA (Rohde \& Schwarz ZVA-50), with each microstrip connection attached to a VNA port.

\section*{Data availability}

Source data are provided with this paper.

\section*{Code availability}

The source code for the PNGF method is publicly available at: \url{https://github.com/ACME-Lab-Stanford/PNGF} \cite{sun_pngf_2026}.


\section*{Acknowledgements}

The authors gratefully acknowledge support by the Air Force Office of Scientific Research (FA9550-20-1-0087, C.S., and FA9550-25-1-0020, C.S.) and the National Science Foundation (CCF-2047433, C.S.).

\section*{Author information}

\subsection*{Contributions}
C.S. conceived the idea and supervised the work. 
J.H.S. and Y.Z. performed numerical simulations. 
C.S., J.H.S., and Y.Z. carried out the inverse design of the example studies. 
M.E. measured the fabricated devices. 
H.C.L. and C.W.H. implemented augmented partial factorization for precomputations.
J.H.S. and C.S. participated in the writing of this manuscript.

\subsection*{Corresponding author}
Correspondence to Constantine Sideris.

\section*{Ethics declarations}
\subsection*{Competing interests}

The authors declare no competing interests.

\newpage

\renewcommand{\thesection}{SN} 
\renewcommand{\theequation}{S\arabic{equation}} 
\renewcommand{\theHequation}{S\arabic{equation}}
\renewcommand{\thefigure}{SF\arabic{figure}}
\renewcommand{\thetable}{ST\arabic{figure}}
\setcounter{section}{0}
\setcounter{subsection}{0}
\setcounter{equation}{0}
\setcounter{figure}{0}
\setcounter{table}{0}
\setcounter{page}{1}

\begin{center}
    \Large{Supplementary Information}
\end{center}
\vspace{20pt}

\setcounter{tocdepth}{-1}

\makeatletter
\renewcommand{\l@subsection}{\@dottedtocline{2}{1.5em}{3em}}
\makeatother
\makeatletter
\renewcommand{\l@subsubsection}{\@dottedtocline{3}{3em}{4em}}
\makeatother

\addtocontents{toc}{\setcounter{tocdepth}{3}}
\tableofcontents
\vspace{20pt}

\setcounter{tocdepth}{3}
\setcounter{secnumdepth}{3}

\renewcommand{\thesubsection}{\arabic{subsection}}
\renewcommand{\thesubsubsection}{%
\thesubsection.\arabic{subsubsection}}

\renewcommand{\figurename}{Supplementary Fig.}
\renewcommand{\tablename}{Supplementary Table}

\newpage
\section*{Supplementary Figures}

\begin{figure}[!ht]
\centering
\includegraphics[width=\textwidth,trim=10 10 10 10,clip]{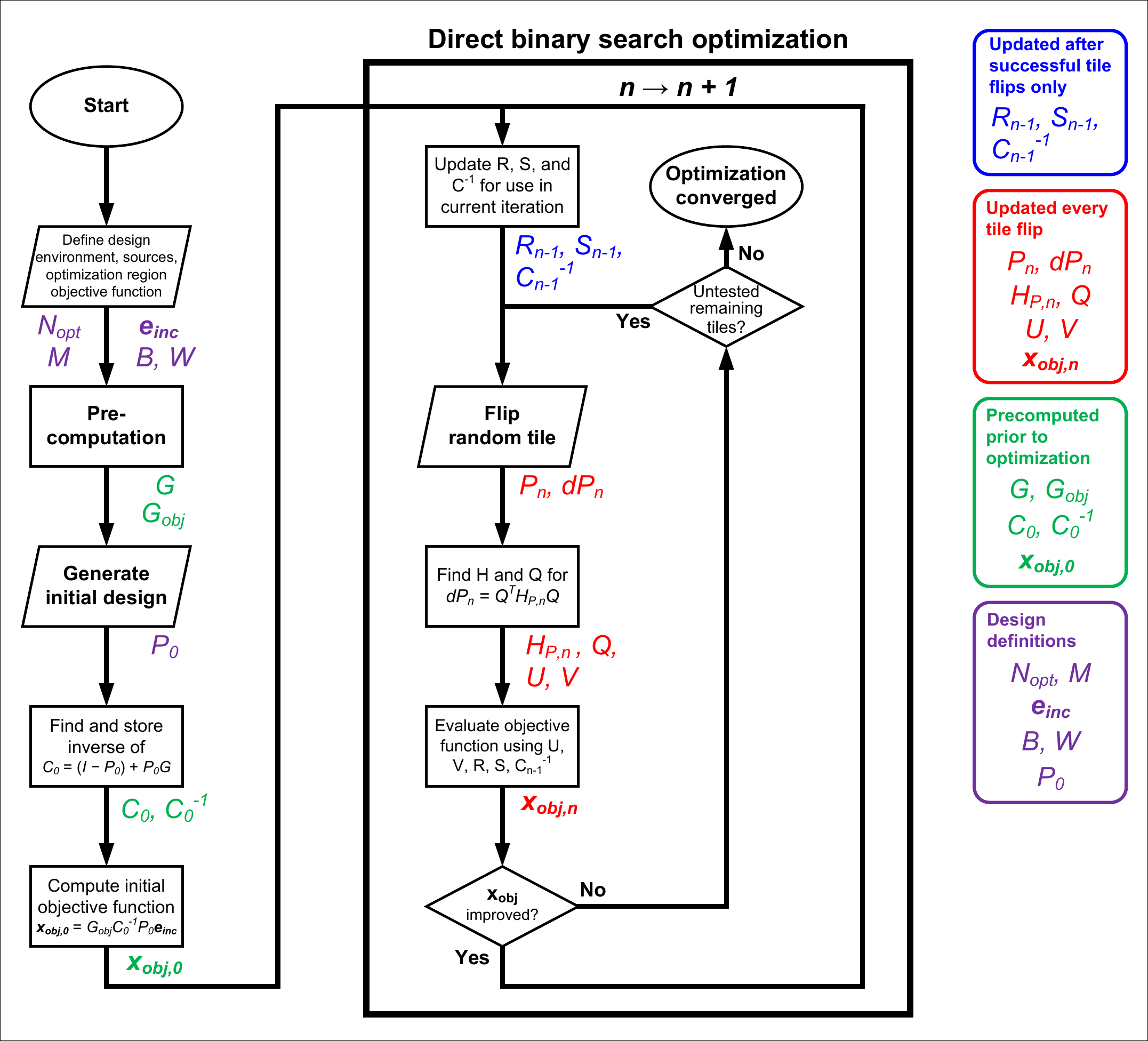}
\caption{\textbf{Flowchart of inverse design utilizing the Precomputed Numerical Green Function method with direct binary search optimization.} Quantities that are obtained in each step are indicated, illustrating the reduction in the computation required for optimization via precomputation and the low-rank update technique.}
\addcontentsline{toc}{section}{\textbf{Supplementary Figure 1}}
\label{fig:supp_flowchart}
\end{figure}

\newpage

\section*{Supplementary Tables}

\addcontentsline{toc}{section}{\textbf{Supplementary Table 1}}

\begin{table}[ht]
\caption{List of symbols}
\label{tb:supp_listofsymbols}
\footnotesize
\begin{tabular}{p{2.1cm} p{10.1cm} }
    \toprule
    \toprule
    $A$ & FDFD Maxwell operator matrix \\ 
    $B$ & projection matrix mapping discretized quantities (e.g., Yee grid spatial locations) in the optimization region to their corresponding indices in the full simulation environment matrix, with size $N_{sim}$ rows by $N_{opt}$ columns \\
    $C$ & PNGF system matrix, square with size $N_{opt}$ \\
    $D^E$, $D^H$ & discretized curl operators of the E and H field, using central finite differences \\
    $D(\theta, \phi)$ & directivity at spherical coordinate polar angle $\theta$ and azimuthal angle $\phi$ \\
    $dC$ & change to $C$ at a given iteration of optimization, diagonal with size $N_{opt}$ \\
    $dP$ & diagonal matrix of size $N_{opt}$ encoding change to $P$ at a given optimization iteration\\
    $\mathbf{E}$ & electric field \\ 
    $\mathbf{e}$ & discretized electric field \\
    $\mathbf{e_{inc}}$ & discretized incident (excitation) electric field over optimization region \\
    $f_{obj}$ & objective function \\
    $G$ & numerical Green function matrix \\
    $G_{obj}$ & matrix that maps the current densities in the optimization region to the objective function vector $\mathbf{x_{obj}}$\\
    $\overline{\overline{G_0}}(\mathbf{r},\mathbf{r'})$ & Free space dyadic Green's function for target point $\mathbf{r}$ evaluated over source points $\mathbf{r'}$ \\
    $\mathbf{H}$ & magnetic field \\
    $H_P$ & diagonal matrix of size $M$ with the nonzero elements of $dP$ \\
    $I$ & identity matrix \\
    $i$ & imaginary unit \\
    $\mathbf{J}$ & current density \\
    $\mathbf{j}$ & discretized current density\\
    $K$ & augmented sparse matrix; partially factorized to find $G$ \\ 
    $K_{obj}$ & augmented sparse matrix; partially factorized to find both $G$ and $G_{obj}$\\
    $K_{L11}$, $K_{U11}$, etc. & block matrices in augmented system \\
    $K_{U22}$ & Schur complement\\ 
    $M$ & number of degrees of freedom in optimization region (e.g., Yee grid spatial locations with finite-difference methods) modified per iteration of optimization\\
    $N_{obj}$ & number of elements in objective function vector $\mathbf{x_{obj}}$\\
    $N_{opt}$ & size of optimization region (i.e. degrees of freedom) \\
    $N_{sim}$ & size of full simulation environment \\
    $P$ & diagonal matrix of size $N_{opt}$ encoding which optimization region locations are filled (1) and which are empty (0) \\
    $p(\mathbf{r})$ & indicator, over continuous space, of whether point $\mathbf{r}$ is metal or free space \\
    $Q$ & projection matrix such that $dP = Q^T H_{p} Q$ \\
    $R$ & matrix used for efficient $\mathbf{x_{obj}}$ evaluation, with size $N_{obj}$ rows by $N_{opt}$ columns \\
    $S$ & vector used for efficient $\mathbf{x_{obj}}$ evaluation, with $N_{opt}$ elements \\
    $S_{11}$, $S_{21}$, etc. & scattering parameters \\
    $U$ & matrix in Woodbury identity, with size $N_{opt}$ rows by $M$ columns \\
    $V$ & matrix in Woodbury identity, with size $M$ rows by $N_{opt}$ columns \\
    $W$ & projection matrix, defined such that $W^T$ maps the field over the full simulation environment to $\mathbf{x_{obj}}$ \\
    $\mathbf{x_{inc}}$ & contribution of $\mathbf{e_{inc}}$ to $\mathbf{x_{obj}}$; given by $W^T \mathbf{e_{inc}}$ \\
    $\mathbf{x_{obj}}$ & vector of quantities needed to compute the objective function \\
    $Y$ & generalized design matrix for non-PEC materials in the optimization region \\
    $\Phi$ & levelset function \\
    $\varepsilon$ & permittivity \\
    $\mu$ & permeability \\
    $\sigma$ & conductivity \\
    $\bm{\tau}$ & vector of parameters defining $\Phi$ \\
    $\omega$ & frequency \\
    \toprule
    \toprule
\end{tabular}
\end{table}

\section*{Supplementary Notes}
\addcontentsline{toc}{section}{\textbf{Supplementary Notes}}

\subsection{Current equivalence derivation}
\label{supp:ce_deriv}

The electric field-only, source-free Maxwell's equations are
\begin{equation}
\omega^2\varepsilon\mathbf{E} - \nabla \times \mu^{-1} \nabla \times \mathbf{E} =  
0.
\end{equation}
Assuming no magnetic materials (these can be incorporated with additional magnetic polarization densities but are not relevant to the problems considered in this work), $\mu = \mu_0$. We can introduce an equivalent polarization density $\mathbf{J}_p(\mathbf{r}) = i\omega\varepsilon_0(\varepsilon_r(\mathbf{r})-1)\mathbf{E}(\mathbf{r})$ to express the total electric fields $\mathbf{E}$ in the presence of an inhomogeneous dielectric volume $\varepsilon_r(\mathbf{r})$ and rewrite in terms of the free-space Maxwell's equations:
\begin{equation}
\omega^2 \varepsilon_0 \mu_0\mathbf{E} - \nabla \times \nabla \times \mathbf{E} =  
i\omega\mu_0\mathbf{J_p}
.
\end{equation}
We seek to find $\mathbf{J_p}$ to produce the same $\mathbf{E}$ in response to an incident field $\mathbf{E_{inc}}$ in free space, as 
scattered from the material(s) in the optimization region
of a candidate design. 
The excitation $E_{inc}$ may be provided by, e.g., a localized current source or a plane wave.
In free space, the total electric field due to the field produced by a volume electric current density $\mathbf{J}$ and incident field $\mathbf{E_{inc}}$ is given by
\begin{equation}
\mathbf{E} = \mathbf{E_{inc}} - \int_V 
\overline{\overline{G_0}}(\mathbf{r},\mathbf{r'})
\mathbf{J}(\mathbf{r'})
\;dV'
,
\end{equation}
where $\overline{\overline{G_0}}(\mathbf{r},\mathbf{r'})$ is the dyadic free-space Green's function. Substituting in $\mathbf{J_p}$ and multiplying both sides of the equation by $i\omega\varepsilon_0(\varepsilon_r(\mathbf{r})-1)$ yields
\begin{equation}
i\omega\varepsilon_0(\varepsilon_r(\mathbf{r})-1)\mathbf{E} = i\omega\varepsilon_0(\varepsilon_r(\mathbf{r})-1)\mathbf{E_{inc}} - i\omega\varepsilon_0(\varepsilon_r(\mathbf{r})-1)\int_V \overline{\overline{G_0}}(\mathbf{r},\mathbf{r'})
\mathbf{J_p}
\;dV'
.
\end{equation}
By using $\mathbf{J_p}(\mathbf{r}) = i\omega\varepsilon_0 (\varepsilon_r(\mathbf{r})-1)\mathbf{E}(\mathbf{r})$, this can be rewritten as:
\begin{equation}
\mathbf{J_p}(\mathbf{r})=i\omega\varepsilon_0(\varepsilon_r(\mathbf{r})-1) \mathbf{E_{inc}} - 
i\omega\varepsilon_0(\varepsilon_r(\mathbf{r})-1)\int_V \overline{\overline{G_0}}(\mathbf{r},\mathbf{r'})\mathbf{J_p}
\;dV'
.
\label{eq:supp_ce_evie_j}
\end{equation}
The permittivity $\varepsilon_r(\mathbf{r})$ for a metal may be represented as $\varepsilon_r(\mathbf{r}) = 1 + \frac{\sigma(\mathbf{r})}{i\omega\varepsilon_0}$, where $\sigma(\mathbf{r})$ is the material conductivity. Thus,
\begin{equation}
\mathbf{J_p}(\mathbf{r}) = 
\sigma(\mathbf{r})\mathbf{E}_{inc} 
- \sigma(\mathbf{r})\int_V \overline{\overline{G_0}}(\mathbf{r},\mathbf{r'})\mathbf{J_p} 
\; dV'
.
\label{eq:supp_ce_evie_sigma}
\end{equation}
For optimization regions that comprise only metallic object(s) and the background medium (e.g., in this derivation, free space), 
$\sigma$ is either 0 (background) or $\infty$ (metal), where metal is represented by perfect electrical conductor (PEC). As such, by introducing the auxiliary quantity $p(\mathbf{r}) = \frac{\sigma(\mathbf{r})}{1 + \sigma(\mathbf{r})}$ and therefore $\sigma(\mathbf{r})=\frac{p(\mathbf{r})}{1-p(\mathbf{r})}$, equation (\ref{eq:currequiv_j_cont}) in the main text is obtained and reproduced here:
\begin{equation}
p(\mathbf{r})\mathbf{E}_{inc} = (1-p(\mathbf{r}))\mathbf{J_p}(\mathbf{r}) + p(\mathbf{r})\int_V \overline{\overline{G_0}}(\mathbf{r},\mathbf{r'})\mathbf{J_p} 
\; dV'
.
\end{equation}
Note that the variable $p(\mathbf{r})$ was introduced such that $p=0$ corresponds to $\sigma=0$ (free-space) and $p=1$ corresponds to $\sigma=\infty$ (PEC) so that the resulting numerical system can express both free-space and PEC with finite quantities. This resulting integral equation can be discretized using a suitable method of choice as discussed in the main text, and the dyadic free space Green's function $\overline{\overline{G_0}}$ may be replaced with a numerically-computed Green function to incorporate any background environment (arbitrary materials, metals, etc.) in the simulation domain.

\newpage
\subsection{Finite-difference formulation for augmented partial factorization}
\label{supp:fdfd_apf}

The finite-difference frequency domain (FDFD) linear system, which is a frequency-domain discretization of Maxwell's equations, is given by
\begin{align}
D^E \mathbf{E} &= -i\omega\mathrm{diag}(\mathbf{\mu}) \mathbf{H} , \\
D^H \mathbf{H} &= i\omega\mathrm{diag}(\bm{\varepsilon}) \mathbf{E} + \mathbf{J}   
,
\end{align}
where $D^E$ and $D^H$ are matrices that discretize the curl operator using central finite differences, $\mathbf{E}$ and $\mathbf{H}$ are the electric and magnetic field, $\mathbf{J}$ is the current density, $\text{diag}(\varepsilon)$ and $\text{diag}(\mu)$ are diagonal matrices whose entries are the permittivity and permeability, respectively, at each point in the discretized simulation domain, $\omega$ is the frequency, and $i$ is the imaginary unit. The magnetic field $\mathbf{H}$ may be eliminated, yielding
\begin{align}
\left[-i\omega\mathrm{diag}(\bm{\varepsilon}) 
+ \frac{i}{\omega}D^H \mathrm{diag}(\mathbf{\mu}^{-1})D^E\right] \mathbf{E} 
&=  \mathbf{J}    
.
\end{align}
The matrix $\left[-i\omega\mathrm{diag}(\bm{\varepsilon}) 
+ \frac{i}{\omega}D^H \mathrm{diag}(\mathbf{\mu}^{-1})D^E\right]$ is the FDFD system matrix $A$, and solving the linear system $A\mathbf{E} = \mathbf{J}$ by inverting $A$ yields the electric fields produced due to time-harmonic sources represented by $\mathbf{J}$. 

The matrix $B^T A^{-1} B$ in equation (\ref{eq:precomp_syseq_dom}), which is the $G$ matrix considered in precomputation, corresponds to
\begin{equation}
B^T A^{-1} B = 
B^T i \omega \left[\omega^2\mathrm{diag}(\bm{\varepsilon}) - D^H \mathrm{diag}(\mathbf{\mu}^{-1})D^E\right]^{-1} B
\end{equation}
and may be computed with a sparse direct solver such as APF or any linear system solver of choice, as discussed in the Results section (Precomputation) of the main text. For multi-frequency optimization, multiple $A$ matrices may be set up for each frequency $\omega$ of interest and factorized to obtain $G$ matrices for each $\omega$.

\newpage
\subsection{Cost of system matrix update and objective function evaluation}
\label{supp:pngfcost}

For evaluating the objective function $\mathbf{x_{obj, n}}$ after each attempted tile flip using equation (\ref{eq:lru_eobj_rs}), the product 
$R_{n-1} dP_n \mathbf{e_{inc}}$
requires $(N_{obj} + 1)M$ floating-point scalar multiplication operations to perform, where $N_{obj}$ corresponds to the number of elements in $\mathbf{x_{obj,n}}$ and $M$ is the number of nonzero entries in the $dP_n$ diagonal matrix. 
We now consider the product
\begin{equation}
R_{n-1} U \left( I + V C_{n-1}^{-1} U \right)^{-1} V \left(C_{n-1}^{-1} dP_n \mathbf{e_{inc}} + S_{n-1}\right)
.
\label{eq:supp_objeval_prod_ruivcuvcpe}
\end{equation}
The $C_{n-1}^{-1} dP_n \mathbf{e_{inc}} + S_{n-1}$ term requires $M(N_{opt} + 1)$ operations, not including adding the vector $S_{n-1}$ to the product. 
The multiplication $V C_{n-1}^{-1} U$ may take up to the order of $M^2 N_{opt} + M N_{opt}$ operations to perform. 
Carrying out the matrix inversion for $\left(I + V C_{n-1}^{-1} U\right)$ requires in general $\mathcal{O}(M^3)$ operations. 
Once these quantities have been obtained, the product (\ref{eq:supp_objeval_prod_ruivcuvcpe}) is a multiplication of matrices of sizes of, from left to right, $(N_{obj} \times N_{opt})$, $(N_{opt} \times M)$ (sparse), $(M \times M)$, $(M \times N_{opt})$, and $(N_{opt} \times 1)$. Performing the multiplication from right to left requires 
$MN_{opt} + M^2 + MN_{obj} + M$
operations, when the sparsity of $U$ is taken into account. 
Since $M \ll N_{opt}$ and $N_{obj}\ll N_{opt}$, the first term dominates the required number of operations for (\ref{eq:supp_objeval_prod_ruivcuvcpe}), including the inversion of $\left(I + V C_{n-1}^{-1} U\right)$.
As such, the cost of finding the objective function is 
$\mathcal{O}((M^2 + 3M)N_{opt})$
which is $\mathcal{O}(N_{opt})$ with respect to the size $N_{opt}$ of the optimization region.

For updates to the PNGF system matrix $C$ using equation (\ref{eq:lru_woodbury}), 
the product 
\begin{equation}
C_{n-1}^{-1} U \left( I + V C_{n-1}^{-1} U\right)^{-1} V C_{n-1}^{-1}
\label{eq:supp_ceval_prod}
\end{equation}
is a multiplication of matrices with sizes of, from left to right, $(N_{opt} \times N_{opt})$, $(N_{opt} \times M)$ (sparse), $(M \times M)$, $(M \times N_{opt})$, and $(N_{opt} \times N_{opt})$. The quantities $C_{n-1}^{-1} U$ and $\left( I + V C_{n-1}^{-1} U\right)^{-1}$ were found when the objective function was computed (prior to deciding to retain the tile flip and to update the system matrix). When performed in the appropriate order,
the number of multiplication operations involved in the product (\ref{eq:supp_ceval_prod}) is 
$2M N_{opt}^2 + M^2N_{opt}$.
As with computing the objective function, the first term dominates the required number of operations for the product. 
Therefore, the overall cost of updating the system matrix is $\mathcal{O}(M N_{opt}^2)$, which is $\mathcal{O}(N_{opt}^2)$ with respect to $N_{opt}$.

\newpage
\subsection{Generalized optimization region materials}
\label{supp:pngf_curreq_ext}

The use of conductivity in the derivation of the PNGF linear system, in equation (\ref{eq:supp_ce_evie_sigma}), and the change of variables with $p$ yields a system for an optimization region where each discretized unknown, as encoded in the matrix $P$, is binary, either empty or perfect electrical conductor (PEC). However, the PNGF method is not restricted to this, and other materials can be represented in the optimization region. 
Here, it is important to note that PNGF is also not limited to pixelated devices. While the method has been developed in the main text using such designs with DBS to facilitate fabrication of the example designs, updates to the PNGF system can modify any number of unknowns in the optimization region, and the concept of a pixel or tile is not intrinsic to the method. 

We define
\begin{align}
\begin{split}
Y &=
i \omega \varepsilon_0 \left(
\begin{bmatrix}
\varepsilon_{r,1}^N & 0 & \hdots & 0 \\
0 & \varepsilon_{r,2}^N & \hdots & 0 \\
\vdots & \vdots & \ddots & \vdots \\
0 & 0 & \hdots & \varepsilon_{r,N_{opt}}^N\\
\end{bmatrix}
- \begin{bmatrix}
\varepsilon_{r,1}^O & 0 & \hdots & 0 \\
0 & \varepsilon_{r,2}^O & \hdots & 0 \\
\vdots & \vdots & \ddots & \vdots \\
0 & 0 & \hdots & \varepsilon_{r,N_{opt}}^O\\
\end{bmatrix}
\right) \nonumber\\
&=
\begin{bmatrix}
i \omega \varepsilon_0(\varepsilon_{r,1}^N - \varepsilon_{r,1}^O) & 0 & \hdots & 0 \\
0 & i \omega \varepsilon_0(\varepsilon_{r,2}^N - \varepsilon_{r,2}^O) & \hdots & 0 \\
\vdots & \vdots & \ddots & \vdots \\
0 & 0 & \hdots & i \omega \varepsilon_0(\varepsilon_{r,N_{opt}}^N - \varepsilon_{r,N_{opt}}^O) 
\end{bmatrix}
.
\label{eq:supp_ormat_generic}
\end{split}
\end{align}
Each entry in this discretized diagonal matrix corresponds to the difference between the new permittivity of the optimization region, 
as is being encoded by $Y$,
at the given discretized spatial location (e.g., for finite-difference discretization, a Yee cell edge)
and the original permittivity in the optimization region as represented by the numerical Green function, $G$. The entry $\varepsilon_{r, m}^N$ at the $m$th index of the diagonal of $Y$ is the complex-valued relative permittivity at the corresponding discretized spatial location; that is,
\begin{equation}
\varepsilon_{r, m}^N = \varepsilon_{r, m}' + 
i\varepsilon_{r, m}'' = 1 + \chi_{m}
,
\end{equation}
where $\varepsilon_{r, m}'$ is the real part and $\varepsilon_{r, m}''$ is the imaginary part of the relative permittivity, and $\chi_{m}$ is the susceptiblity.
If the $m$th index represents metal, then, at RF or microwave frequencies, we may write
\begin{equation}
\varepsilon_{r, m} = 1 + \frac{\sigma_m}{i \omega \varepsilon_0}
,
\end{equation}
and the $m$th element of $Y$ is simply $\sigma_m$ (assuming the background medium $\varepsilon_{r,m}^O$ is $\varepsilon_0$). 
Note that for the PEC case, since $\sigma\to\infty$, the corresponding entry in $Y\to\sigma$, even when the background medium $\varepsilon_r^O$ is not $\varepsilon_0$.

Discretizing equation (\ref{eq:supp_ce_evie_j}) using $Y$ yields
\begin{equation}
(I + YG) \mathbf{j} = Y \mathbf{e_{inc}}
,
\label{eq:supp_ce_discr}
\end{equation}
where, as with equation (\ref{eq:currequiv_j_disc}) in the main text, $\mathbf{j}$ and $\mathbf{e_\text{inc}}$ are the discretized polarization density and incident electric field vectors over the optimization region. The matrix encoding the design has been generalized to represent generic media in the optimization region. The numerical Green's function matrix $G$ is identical to that for the binary PEC current-equivalence formulation, provided that the simulation environment is the same. It is apparent that by multiplying equation (\ref{eq:supp_ce_discr}) by $(I-P)$ and substituting in the identity $(I-P)Y=P$, equation (\ref{eq:supp_ce_discr}) becomes exactly equation (\ref{eq:currequiv_j_disc}) in the main text for the specialized PEC case. The system matrix $C$ is now given by 
\begin{equation}
    C = I + YG
    .
\end{equation}

In general, the permittivities $\varepsilon_{r, m}^N$ are frequency-dependent.
For optimization with such materials, at each frequency of optimization, numerical Green function matrices may be obtained and a linear system (i.e., equation (\ref{eq:supp_ce_discr})) may be set up. With the binary PEC formulation, the design matrix $P$ is independent of frequency and is the same for the multiple systems (each corresponding to one frequency of optimization). However, with this generalized case, multiple $Y$ matrices, one for each frequency of optimization, will be present. Each matrix is determined by the frequency-dependent behavior of each $\varepsilon_{r, m}^N$, modeling a given material, in the optimization region.

Any suitable optimization algorithm that can apply sparse updates to the design region at each iteration, such as DBS, genetic algorithms, and particle swarm optimization, can be used to optimize the design $Y$ using the low-rank update methods discussed in the Results section (Low-rank update evaluation) of the main text. As such, the efficiency and complexity of the PNGF method remain exactly the same for the generalized material case.
Suppose that the chosen algorithm modifies $M$ elements in the design matrix $Y_n$ in the optimization region at the $n$th iteration of optimization. Let $dY_n = Y_n - Y_{n-1}$ represent the change to $Y$. As with the binary PEC formulation, $dY_n = Q^T H_{Y,n} Q$ may be defined, where $H_{Y,n}$ is a diagonal $M$-by-$M$ matrix whose entries are the nonzero elements of $dY_n$. Then, with $U = Q^T H_{Y,n}$ and $V = QG$, the update is given by
\begin{equation}
dC_n = dY_n G = \left[Q^T H_{Y,n}\right]\left[QG \right] = UV
.
\label{eq:supp_lru_dcn_generalized}
\end{equation} 
The system matrix may be updated and the objective function, defined as 
\begin{equation}
    \mathbf{x_{obj}} = G_{obj}C^{-1}Y\mathbf{e_{inc}} + \mathbf{x_{inc}}
    ,
    \label{eq:supp_xobj_y}
\end{equation}
may be evaluated efficiently using the same technique as with equations (\ref{eq:lru_woodbury}) and (\ref{eq:lru_eobj_rs}) in the main text.

It is also possible to reformulate the PNGF linear system in terms of $\mathbf{e_{opt}} = G\mathbf{j_{opt}}$, the fields in the optimization region corresponding to a solution $\mathbf{j_{opt}}$. 
While this does not reduce the complexity of evaluating $\mathbf{x_{obj}}$ or updating $C^{-1}$, this enables the gradient of the objective function with respect to all $N_{opt}$ design parameters in $Y$ to be obtained with $\mathcal{O}(N_{opt})$ complexity, as detailed in Supplementary Note \ref{supp:gradient}. 
With $\widetilde{C} = G^{-1} + Y$ and $\widetilde{G_{obj}} = G_{obj}G^{-1}$, equation (\ref{eq:supp_ce_discr}) becomes 
\begin{equation}
    (G^{-1} + Y)\mathbf{e_{opt}} = 
    \widetilde{C}\mathbf{e_{opt}} = 
    Y \mathbf{e_{inc}}
    ,
\end{equation}
and equation (\ref{eq:supp_xobj_y}) becomes
\begin{equation}
    \mathbf{x_{obj}} = \widetilde{G_{obj}} \widetilde{C}^{-1}Y\mathbf{e_{inc}} + \mathbf{x_{inc}}
    .
    \label{eq:supp_xobj_y_tilde}
\end{equation}
Woodbury updates to $\widetilde{C}^{-1}$ may be performed with 
$U = Q^T H_{Y,n}$ and $V = Q$, where the update is given by
\begin{equation}
d\widetilde{C}_n = dY_n = \left[Q^T H_{Y,n}\right]\left[Q \right] = UV
.
\label{eq:supp_lru_dcn_generalized_tilde}
\end{equation} 
Note that with this alternate $\mathbf{e_{opt}}$ based formulation, both the $U$ and $V$ matrices are sparse.

\newpage
\subsection{Gradient-based optimization}
\label{supp:gradient}

With the generalized optimization region formulation introduced in Supplementary Note \ref{supp:pngf_curreq_ext}, the gradient of the objective function with respect to the design parameters $\bm{\varepsilon_r}$ over the optimization region may be computed directly without any EM simulations. We show that obtaining the gradient over the full optimization region, with $N_{opt}$ degrees of freedom, has a complexity of only $\mathcal{O}(N_{opt})$.
As such, gradient-based optimization may be leveraged efficiently with PNGF for dielectric problems. 

\subsubsection{Objective function gradient}

The objective function $f_{obj}$ is a real-valued scalar function which takes the vector of objective function quantities $\mathbf{x_{obj}}$, which is in general complex-valued, as input. Let us assume that the input permittivities $\bm{\varepsilon_r}$ are purely real, and let us define a vector $\mathbf{y} = i \omega \varepsilon_0 (\bm{\varepsilon_r^N}-\bm{\varepsilon_r^O}) = i \omega \varepsilon_0 \bm{\Delta\varepsilon_r} $, so the matrix $Y = \text{diag}(\mathbf{y})$. We will evaluate the gradient by using the chain rule, evaluating the derivatives of $f_{obj}$ with respect to $\mathbf{x_{obj}}$, of $\mathbf{x_{obj}}$ with respect to $\mathbf{y}$, and of $\mathbf{y}$ with respect to $\bm{\Delta\varepsilon_r}$. Since $\mathbf{x_{obj}}$ and $\mathbf{y}$ are complex quantities, Wirtinger derivatives must be considered. However, noting that $f_{obj}$ is a real-valued function and that $\mathbf{y}$ is purely imaginary, the chain rule simplifies to
\begin{equation}
    \nabla f_{obj} 
    =
    \frac{d f_{obj}}{d \bm{\varepsilon_r}}
    =
    2
    \text{Re}
    \left[ 
    \frac{\partial f_{obj}}{\partial \mathbf{x_{obj}}} 
    \frac{\partial \mathbf{x_{obj}}}{\partial \mathbf{y}}  
    \frac{d \mathbf{y}}{d \bm{\varepsilon_r}} 
    \right]
    ,
    \label{eq:supp_grad_chainrule}
\end{equation}
where $\frac{\partial f_{obj}}{\partial \mathbf{x_{obj}}}$ is the Wirtinger derivative of $f_{obj}$ with respect to $\mathbf{x_{obj}}$ and $\frac{\partial \mathbf{x_{obj}}}{\partial \mathbf{y}}$ is the Wirtinger derivative of $\mathbf{x_{obj}}$ with respect to $\mathbf{y}$. 

The derivative $\frac{\partial f_{obj}}{\partial \mathbf{x_{obj}}}$ is dependent on the definition of $f_{obj}(\mathbf{x_{obj}})$, and is assumed to be determined prior to optimization. An example objective function is
\begin{equation}
    f_{obj}(\mathbf{x_{obj}}) 
= ||\mathbf{x_{obj}}||^2
= \sum_{m = 1}^{N_{opt}} |x_{obj,m}|^2
= \sum_{m = 1}^{N_{opt}} x_{obj,m} \overline{x_{obj,m}},
,
\end{equation}
the sum of squared magnitudes of the objective function vector quantities. In this case, the Wirtinger derivative with respect to $\mathbf{x_{obj}}$ is $\frac{\partial f_{obj}}{\partial \mathbf{x_{obj}}} =  \overline{\mathbf{x_{obj}}}$. The derivative $\frac{d \mathbf{y}}{d \bm{\varepsilon_r}}$ is given by
\begin{equation}
    \frac{d \mathbf{y}}{d \bm{\varepsilon_r}}
    =
    i \omega \varepsilon_0 I
    .
\end{equation}

Now we find $\frac{\partial \mathbf{x_{obj}}}{\partial \mathbf{y}}$. We employ the $C \rightarrow \widetilde{C}$ formulation discussed in Supplementary Note \ref{supp:pngf_curreq_ext}.
By the product rule,
\begin{equation}
\frac{\partial \mathbf{x_{obj}}}{\partial \mathbf{y}}
= 
\widetilde{G_{obj}} 
\left[
\left(\frac{\partial \widetilde{C}^{-1}}{\partial \mathbf{y}}\right) Y +  \widetilde{C}^{-1} \left(\frac{\partial Y}{\partial \mathbf{y}}\right)
\right]
\mathbf{e_{inc}}
.
\label{eq:supp_grad_pxobj_py}
\end{equation}
We recall the identity for the derivative of the inverse of a matrix $A(p)$ with respect to a scalar or vector $p$:
\begin{equation}
    \frac{d}{dp}A(p) = - A^{-1}(p) \frac{dA}{dp} A^{-1}(p)
    .
\end{equation}
It is useful to consider the evaluation of equation (\ref{eq:supp_grad_pxobj_py}) elementwise, where
\begin{equation}
    \frac{\partial \mathbf{x_{obj}}}{\partial \mathbf{y}}
    =
    \begin{bmatrix}
        \quad &
        \dfrac{\partial \mathbf{x_{obj}}}{\partial y_1} & 
        \quad\dots\quad & 
        \dfrac{\partial \mathbf{x_{obj}}}{\partial y_{N_{opt}}}
        & \quad
    \end{bmatrix}
    .
\end{equation}
For each $m \in [1, N_{opt}]$, the elementwise derivative $\frac{dY}{dy_m}$ is a $(N_{opt} \times N_{opt})$ matrix with a single nonzero element, a 1 at the $(m, m)$th index. This can be thought of as the outer product of the $m$th standard basis vector of length $N_{opt}$ with its transpose. Let us denote this as $X_m$. Additionally, 
\begin{equation}
    \frac{\partial \widetilde{C}^{-1}}{\partial y_m}
    =
    -\widetilde{C}^{-1} \frac{\partial \widetilde{C}}{\partial y_m} \widetilde{C}^{-1}\\
    = 
    -\widetilde{C}^{-1} \frac{dY}{dy_m} \widetilde{C}^{-1}
    .
\end{equation}
So each element $\frac{\partial \mathbf{x_{obj}}}{\partial y_m}$ of the derivative is given by
\begin{align}
    \begin{split}
    \frac{\partial \mathbf{x_{obj}}}{\partial y_m} 
    &=
    \widetilde{G_{obj}} 
    \left(
    -\widetilde{C}^{-1} X_m \widetilde{C}^{-1} Y
    + \widetilde{C}^{-1} X_m 
    \right)
    \mathbf{e_{inc}} \\
    &= 
    \widetilde{G_{obj}}
    \widetilde{C}^{-1} X_m (I - \widetilde{C}^{-1}Y)
    \mathbf{e_{inc}}.
    \label{eq:supp_grad_dxobj_dy_elementwise}
    \end{split}
\end{align}
If we instead consider $\frac{\partial \mathbf{x_{obj}}}{\partial \mathbf{y}}$ directly, the derivative $\frac{d Y}{\partial \mathbf{y}}$ is a third-order tensor of dimensions $(N_{opt} \times N_{opt} \times N_{opt})$, where the $m$th element along the third (last) dimension is a $(N_{opt} \times N_{opt})$ matrix, $X_m$. Let us denote this tensor as $\overline{\overline{T}}$. We can thus write
\begin{equation}
    \frac{\partial \mathbf{x_{obj}}}{\partial \mathbf{y}}
    =
    \widetilde{G_{obj}}
    \widetilde{C}^{-1} \overline{\overline{T}} (I - \widetilde{C}^{-1}Y)
    \cdot
    \mathbf{e_{inc}},
    \label{eq:supp_grad_dxobj_dy_full}
\end{equation}
where the $\cdot$ operation denotes contraction along the second index resulting from the multiplication of a $(N_{opt} \times N_{opt} \times N_{opt})$ with a $(N_{opt} \times 1)$ vector, 
\begin{equation}
    \left[
        \widetilde{G_{obj}}
        \widetilde{C}^{-1} \overline{\overline{T}} (I - \widetilde{C}^{-1}Y)
        \cdot
        \mathbf{e_{inc}}
    \right]_{km}
    =
    \sum_{l=1}^{N_{opt}}
    \left[
        \widetilde{G_{obj}}
        \widetilde{C}^{-1} \overline{\overline{T}} (I - \widetilde{C}^{-1}Y)
    \right]_{klm}
    e_{inc,l}
    .
\end{equation}
However, in practice elementwise evaluation with equation (\ref{eq:supp_grad_dxobj_dy_elementwise}) may be preferable for implementation. 

With the above, the full gradient can be written as
\begin{equation}
    \nabla f_{obj}
    =
    2
    \text{Re}
    \left[ 
    \frac{\partial f_{obj}}{\partial \mathbf{x_{obj}}} 
    \left(
    \widetilde{G_{obj}}
    \widetilde{C}^{-1} \overline{\overline{T}}(I - \widetilde{C}^{-1}Y)
    \cdot
    \mathbf{e_{inc}}
    \right)
    \left( i \omega \varepsilon_0 I \right)
    \right]
    ,
\end{equation}
which may be simplified to
\begin{equation}
    \nabla f_{obj}
    =
    - 2 \omega \varepsilon_0
    \text{Im}
    \left[ 
    \frac{\partial f_{obj}}{\partial \mathbf{x_{obj}}} 
    \widetilde{G_{obj}}
    \widetilde{C}^{-1} \overline{\overline{T}}(I - \widetilde{C}^{-1}Y)
    \cdot
    \mathbf{e_{inc}}
    \right]
    .
    \label{eq:supp_grad_full}
\end{equation}
In elementwise form, this can be written as
\begin{equation}
    \frac{\partial f_{obj}}{\partial \bm{\varepsilon_r}}
    =
    \begin{bmatrix}
        \quad &
        \dfrac{\partial f_{obj}}{\partial \varepsilon_{r,1}} & 
        \quad\dots\quad & 
        \dfrac{\partial f_{obj}}{\partial \varepsilon_{r, N_{opt}}}
        & \quad
    \end{bmatrix}
    ,
\end{equation}
where for $m \in [1, N_{opt}]$,
\begin{equation}
    \dfrac{\partial f_{obj}}{\partial \varepsilon_{r,m}}
    =
    - 2 \omega \varepsilon_0
    \text{Im}
    \left[ 
    \frac{\partial f_{obj}}{\partial \mathbf{x_{obj}}} 
    \widetilde{G_{obj}}
    \widetilde{C}^{-1} X_m(I - \widetilde{C}^{-1}Y)
    \mathbf{e_{inc}}
    \right]
    .
\end{equation}

\subsubsection{Computational efficiency}

In assessing the cost of computing the gradient, we make the following assumptions: 
First, $\frac{\partial f_{obj}}{\partial \mathbf{x_{obj}}}$ is known a priori since this is based on the definition of $f_{obj}$. 
Second, the products $\widetilde{G_{obj}}\widetilde{C}^{-1}$ and $\widetilde{C}^{-1} Y \mathbf{e_{inc}}$ are known from computing $\mathbf{x_{obj}}$ at the previous iteration of optimization. In particular, at a given iteration, we seek to find the gradient to determine how the design specification $\bm{\varepsilon_r^N} - \bm{\varepsilon_r^O}$ should be updated. In the preceding iteration, these quantities would have been found for the current design configuration $Y$ when obtaining $\mathbf{x_{obj}}$. 
It is apparent that $\widetilde{G_{obj}}\widetilde{C}^{-1}$ and $\widetilde{C}^{-1} Y \mathbf{e_{inc}}$ are the quantities $R$ and $S$, corresponding to equations (\ref{eq:lru_r}) and (\ref{eq:lru_s}) in the main text with the $C \rightarrow \widetilde{C}$ formulation.
Third, the product $\left(\widetilde{G_{obj}}\widetilde{C}^{-1}\right) \overline{\overline{T}}$ (or $\left(\widetilde{G_{obj}}\widetilde{C}^{-1}\right) X_m$ in the elementwise form, which is a $(N_{obj} \times N_{opt})$ matrix containing the $m$th column of $\widetilde{G_{obj}}\widetilde{C}^{-1}$ and zeroes elsewhere) does not incur any cost since this is just splitting $\widetilde{G_{obj}}\widetilde{C}^{-1}$ column by column. 
As in Supplementary Note \ref{supp:pngfcost}, we consider the computational complexity by counting the number of multiplication operations.

To evaluate the cost of computing $\frac{\partial \mathbf{x_{obj}}}{\partial \mathbf{y}}$, we consider the elementwise form and write the terms as follows: 
\begin{equation}
    \left( \widetilde{G_{obj}} \widetilde{C}^{-1} X_m \right) \mathbf{e_{inc}}
    -
    \left( \widetilde{G_{obj}} \widetilde{C}^{-1} X_m \right) \left(\widetilde{C}^{-1} Y \mathbf{e_{inc}}\right)
    \label{eq:supp_grad_pxobj_pyk_eval}
\end{equation}
Both terms are a product of a $(N_{obj} \times N_{opt})$ matrix (with a single nonzero column) with a vector of length $N_{opt}$, requiring $N_{obj}$ operations each. As such, evaluating the quantity (\ref{eq:supp_grad_pxobj_pyk_eval}) for each $m \in [1, N_{opt}]$ requires a total of $2N_{obj}N_{opt}$ operations.

For the remainder of the gradient evaluation, the product $\frac{\partial f_{obj}}{\partial \mathbf{x_{obj}}} \frac{\partial \mathbf{x_{obj}}}{\partial \mathbf{y}}$ is a multiplication of matrices of sizes $(1 \times N_{obj})$ and $(N_{obj} \times N_{opt})$, requiring $N_{obj}N_{opt}$ products.
There are an additional $N_{opt}$ operations from multiplying by the scalar $2\omega\varepsilon_0$ after taking the imaginary part. 
Therefore, the total number of multiplication operations involved in computing the gradient is $(3N_{obj} + 1)N_{opt}$.
As such, the cost of finding the gradient is $\mathcal{O}(N_{opt})$.

\newpage
\subsubsection{Optimization considerations}

At each iteration of inverse design, once the gradient is computed in $\mathcal{O}(N_{opt})$ and the modification to $Y$ is determined via the gradient, the system matrix inverse $C^{-1}$ must be updated. 
In general, this is not a sparse update, and inverting the matrix directly, corresponding to solving the linear system of size $N_{opt}$ without the benefits of low-rank update techniques, incurs a $\mathcal{O}(N_{opt}^3)$ cost. 
It should be noted, however, that gradient-based optimization with PNGF nonetheless provides significant acceleration compared to conventional methods with the full simulation environment (e.g., adjoint method), since the problem size is reduced to $N_{opt} \ll N_{sim}$. 
Furthermore, more sophisticated gradient-based algorithms (e.g., levelset methods) may enforce sparsity in the modification to $Y$ and the corresponding change to $C^{-1}$ by limiting the number of elements modified per iteration to be much smaller than $N_{opt}$. 
This would allow $C^{-1}$ to be updated via equation (\ref{eq:lru_woodbury}) with a complexity of $\mathcal{O}(N_{opt}^2)$, as with direct binary search (DBS) optimization.

\newpage
\subsection{Levelset methods}
\label{supp:levelset}

Levelset methods efficiently represent arbitrary regions of constant material within the optimization domain with levelset functions. In one typical case, a levelset function $\Phi(\mathbf{r}, \bm{\tau})$ is a continuous real-valued scalar function such that
\begin{equation}
    \varepsilon_r(\mathbf{r})
    =
    \begin{cases}
        \varepsilon_{r,1} \qquad \Phi(\mathbf{r}, \bm{\tau}) > 0 \\
        \varepsilon_{r,2} \qquad \Phi(\mathbf{r}, \bm{\tau}) < 0 \\
    \end{cases}
    ,
\end{equation}
where $\varepsilon_r(\mathbf{r})$ is the permittivity at a point $\mathbf{r}$, and $\varepsilon_{r,1}$ and $\varepsilon_{r,2}$ are relative permittivity values. The quantity $\bm{\tau}$ denotes a vector of parameters that define the levelset function, which could be, e.g., center points, scale factors, or radii of basis functions. The boundary between permittivity values is at the zero crossing $\Phi = 0$. When the permittivity distribution is discretized via, e.g., a finite-difference method, permittivity averaging is typically employed to allow the boundary to move continuously.  
The interested reader may refer to \cite{osher_level_2006, zhou_level-set_2010, vercruysse_analytical_2019} and other references for more on levelset methods and their application to electromagnetics optimization. A representation of a levelset function and the corresponding optimization region configuration is shown in Supplementary Fig. \ref{fig:supp_levelsets}(a).

\begin{figure}[!ht]
\centering
\includegraphics[width=0.9\textwidth,trim=10 10 10 10,clip]{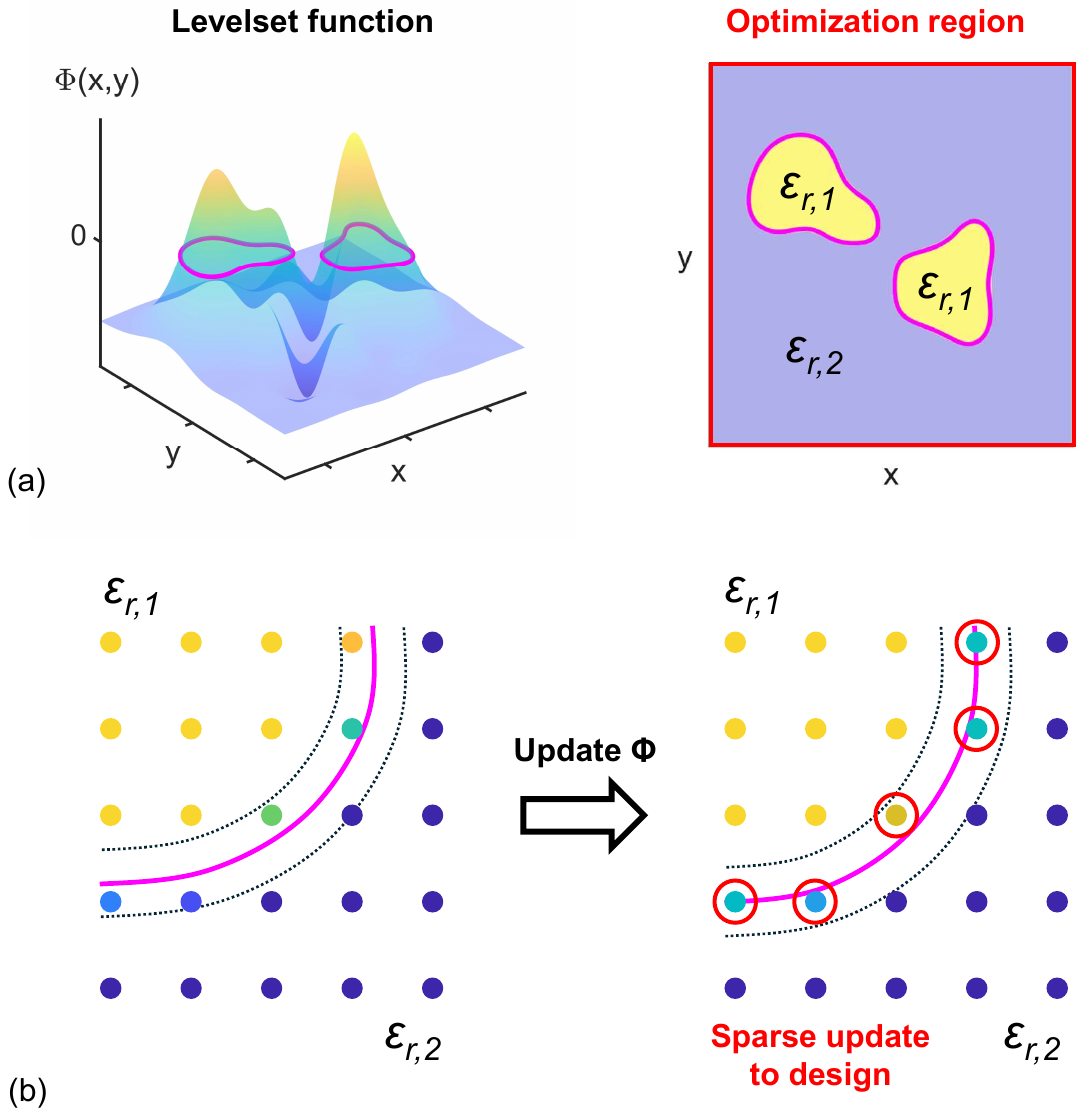}
\caption{
\textbf{Levelset methods.} (a) A representative example levelset function $\Phi$ for a two-dimensional optimization region; (b) Depiction of permittivity averaging for discretized spatial points close to the levelset boundary, and the sparse update to $\bm{\varepsilon_r}$ resulting from the modification to $\Phi$ at a given iteration of optimization. A Cartesian discretization of permittivity is shown in (b) for illustrative purposes and may not be practical for a particular optimization problem of interest.}
\addcontentsline{toc}{subsubsection}{Supplementary Figure 2}
\label{fig:supp_levelsets}
\end{figure}

At the $n$th optimization iteration, $\Phi$ at each spatial location in the optimization region is updated with
\begin{equation}
    \Phi_n = 
    \Phi_{n-1} + \alpha \bm{V_n} |\nabla\Phi_{n-1}|,
\end{equation}
where $\alpha$ is a scalar step size, $\nabla\Phi_{n-1}$ is the gradient of $\Phi_{n-1}$ with respect to the parameters $\bm{\tau}$, and
$\bm{V_n}$ is the normal velocity,
\begin{equation}
    \bm{V_n} = -\frac{df_{obj}}{d\Phi_{n-1}}
    .
\end{equation}
Evolving the levelset function in the direction of the velocity field extremizes $f_{obj}$.

The gradient $\nabla\Phi_{n-1}$ is readily available analytically from $\Phi_{n-1}$. To find $\bm{V}$, the gradient $\frac{d f_{obj}}{d \bm{\varepsilon_r}} = \nabla f_{obj}$ and the derivative $\frac{d \bm{\varepsilon_r}}{d \Phi_{n-1}}$ are required. The derivative $\frac{d \bm{\varepsilon_r}}{d \Phi_{n-1}}$ is also readily obtained from the specific permittivity averaging scheme used. This derivative is a $(N_{opt} \times N_{opt})$ diagonal matrix, as the $m$th element in the permittivity vector $\bm{\varepsilon_r}$ is determined by the value of $\Phi_{n-1}$ at the $m$th discretized degree of freedom in the optimization region. Furthermore, $\frac{d \bm{\varepsilon_r}}{d \Phi_{n-1}}$ is only nonzero for elements close to the boundary. Let $M$ represent the number of elements with nonzero $\frac{d \bm{\varepsilon_r}}{d \Phi_{n-1}}$, with $M  \ll N_{opt}$.

While the gradient $\nabla f_{obj}$ may be computed with equation (\ref{eq:supp_grad_full}), since $\frac{d \bm{\varepsilon_r}}{d \Phi_{n-1}}$ has only $M$ nonzero entries, only the columns of $\nabla f_{obj}$ corresponding to those entries are needed. As such, when computing $\frac{\partial \mathbf{x_{obj}}}{\partial \mathbf{y}}$ with equations (\ref{eq:supp_grad_dxobj_dy_elementwise}-\ref{eq:supp_grad_dxobj_dy_full}), only those $M$ elements need to be included in $Y = \text{diag}(\mathbf{y})$. 
Thus, the dominant $2N_{obj}N_{opt}$ term in the number of operations involved in computing $\frac{\partial \mathbf{x_{obj}}}{\partial \mathbf{y}}$ can be reduced to $2N_{obj}M$.
Additionally, as only the $M$ elements need to be retained in the derivative $\frac{d \mathbf{y}}{d \bm{\varepsilon_r}}$, the overall cost of finding $\nabla f_{obj}$ could be lowered to $\mathcal{O}(M)$.

Next, the product
\begin{equation}
    \frac{df_{obj}}{d\Phi_{n-1}}
    =
    \frac{df_{obj}}{d\bm{\varepsilon_r}}
    \frac{d\bm{\varepsilon_r}}{d\Phi_{n-1}}
\end{equation}
is performed elementwise, requiring $M$ multiplication operations to compute. 
The quantity $\bm{V_n} |\nabla\Phi_{n-1}|$ is the product of a vector of length $N_{opt}$ (with only $M$ nonzero elements) with a $(N_{opt} \times N_{\bm{\tau}})$ matrix, where $N_{\bm{\tau}}$ is the number of elements in $\bm{\tau}$, and this requires $MN_{\bm{\tau}}$ operations. 
Since the number of parameters defining the levelset function may be expected to be far smaller than $N_{opt}$, the update to $\Phi$, and thus the update to $\bm{\varepsilon_r}$, can be computed in $\mathcal{O}(MN_{opt})$ operations.

Once the update to $\bm{\varepsilon_r}$ has been determined for iteration $n$, the new design specification $Y_n$ is used to update the system matrix inverse $\widetilde{C}^{-1}$ and to compute $\mathbf{x_{obj,n}}$. The update $\Delta \Phi_n = \Phi_n - \Phi_{n-1}$ yields a sparse update $dY_n = Y_n - Y_{n-1}$ owing to the small ($M \ll N_{opt}$) number of discretized spatial locations with nonzero $\frac{d \varepsilon_r}{d \Phi_{n-1}}$, as shown in Supplementary Fig. \ref{fig:supp_levelsets}(b). As such, $\widetilde{C}^{-1}$ can be updated by using equation (\ref{eq:supp_lru_dcn_generalized_tilde}) and equation (\ref{eq:lru_woodbury}) in the main text with the change of variables $C \rightarrow \widetilde{C}$, with a cost of $\mathcal{O}(N_{opt}^2)$.

\newpage
\subsection{Non-finite difference discretization}
\label{supp:alt_discr}

Although the examples discussed in the main text were discretized via the finite-difference method, the PNGF method can be used with any suitable discretization approach, including finite element methods (FEM) and method of moments (MoM) integral equations. Here we present a preliminary, proof-of-concept example of utilizing the PNGF method with a MoM solver to rapidly optimize a multilayer RF beamforming metasurface. We employ the commercial Altair FEKO software~\cite{ref_feko} for PNGF precomputation, since FEKO can output the raw MoM interaction matrix, which is needed to extract the numerical Green's function.

An in-depth introduction to MoM and integral equation methods is outside the scope of this Supplementary Note, and the interested reader is encouraged to peruse \cite{volakis_integral_2012, gibson_method_2024} and other references for more background information. FEKO uses MoM and the surface equivalence principle (SEP) to discretize a system, comprising dielectric and metallic objects, with a surface mesh consisting of flat triangular elements. The classical Rao-Wilton-Glisson (RWG)~\cite{rao_electromagnetic_1982} divergence-conforming edge basis functions are used to represent the unknown surface current densities with a single basis function defined for each interior edge of the mesh. The Galerkin method is used to test the equations, yielding a square, full-rank linear system that can be solved for the basis function coefficients that approximate the unknown current densities. FEKO models dielectric objects using the well-known Poggio-Miller-Chang-Harrington-Wu-Tsai (PMCHWT) formulation~\cite{chang_surface_1977}. 

This example considers a dielectric substrate, where the optimization region comprises patterned metal on both sides of the substrate. Due to their very thin nature, the metals are approximated as being infinitesimally thin perfect electrical conductors (PEC) and are modeled using the Electric Field Integral Equation (EFIE). The dielectric substrate requires both surface electric and magnetic current densities to be found, based on the SEP, whereas the PEC optimization region requires only a surface electric current density. 
As such, the following discretized linear system may be constructed
\begin{equation}
    \begin{bmatrix}
        A^\text{o} & A^\text{os} \\
        A^\text{so} & A^\text{s}
    \end{bmatrix}
    \begin{bmatrix}
        \mathbf{x^\text{o}} \\
        \mathbf{x^\text{s}}
    \end{bmatrix}
    =
    \begin{bmatrix}
        \mathbf{b^\text{o}} \\
        \mathbf{b^\text{s}}
    \end{bmatrix}
    ,
    \label{eq:supp_mom_sys1}
\end{equation}
where $\mathbf{x^\text{o}}$ includes the unknowns corresponding to the objects (i.e. metals) in the optimization region, $\mathbf{x^\text{s}}$ includes the unknowns corresponding to the objects (i.e. the substrate) outside the optimization region, $A^\text{o}$ and $A^\text{s}$ encompass scattering interactions within only the objects in the optimization region and the objects outside, respectively, and $A^\text{os}$ and $A^\text{so}$ couple the integral equations together. In this case, $A^s$ is obtained from the PMCHWT formulation for the dielectric substrate outside the optimization region, and $A^o$ is obtained from the EFIE operator. 
Although for this example the optimization region comprises only metal (PEC) and only electric current densities are required, a dielectric formulation such as the PMCHWT may be used instead to optimize dielectric materials in the optimization region.

By eliminating $\mathbf{x^\text{s}}$, the following is obtained
\begin{equation}
    \left[
    A^\text{o} - A^\text{os} \left(A^\text{s}\right)^{-1} A^\text{so}
    \right] \mathbf{x^\text{o}}
    =
    \left[
    \mathbf{b^\text{o}} - A^\text{os} \left(A^\text{s}\right)^{-1} \mathbf{b^\text{s}}
    \right]
    \label{eq:supp_mom_sys2}
\end{equation}
where $A^\text{o} - A^\text{os} \left(A^\text{s}\right)^{-1} A^\text{so}$ is the numerical Green's function matrix $G$ and the right hand side is the excitation $\mathbf{e_{inc}}$. When precomputation is performed, the system (\ref{eq:supp_mom_sys2}) corresponds to a device where the optimization region is completely filled with metal; that is, equation (\ref{eq:supp_mom_sys2}) is
$G \mathbf{j} = P \mathbf{e_{inc}}$
with $P = I$. Once $G$ is obtained, the PNGF system equation (\ref{eq:currequiv_j_disc}) and the low-rank update and objective function evaluation techniques discussed in the main text may be used for optimization.

\begin{figure}[t]
\centering
\includegraphics[width=\textwidth]{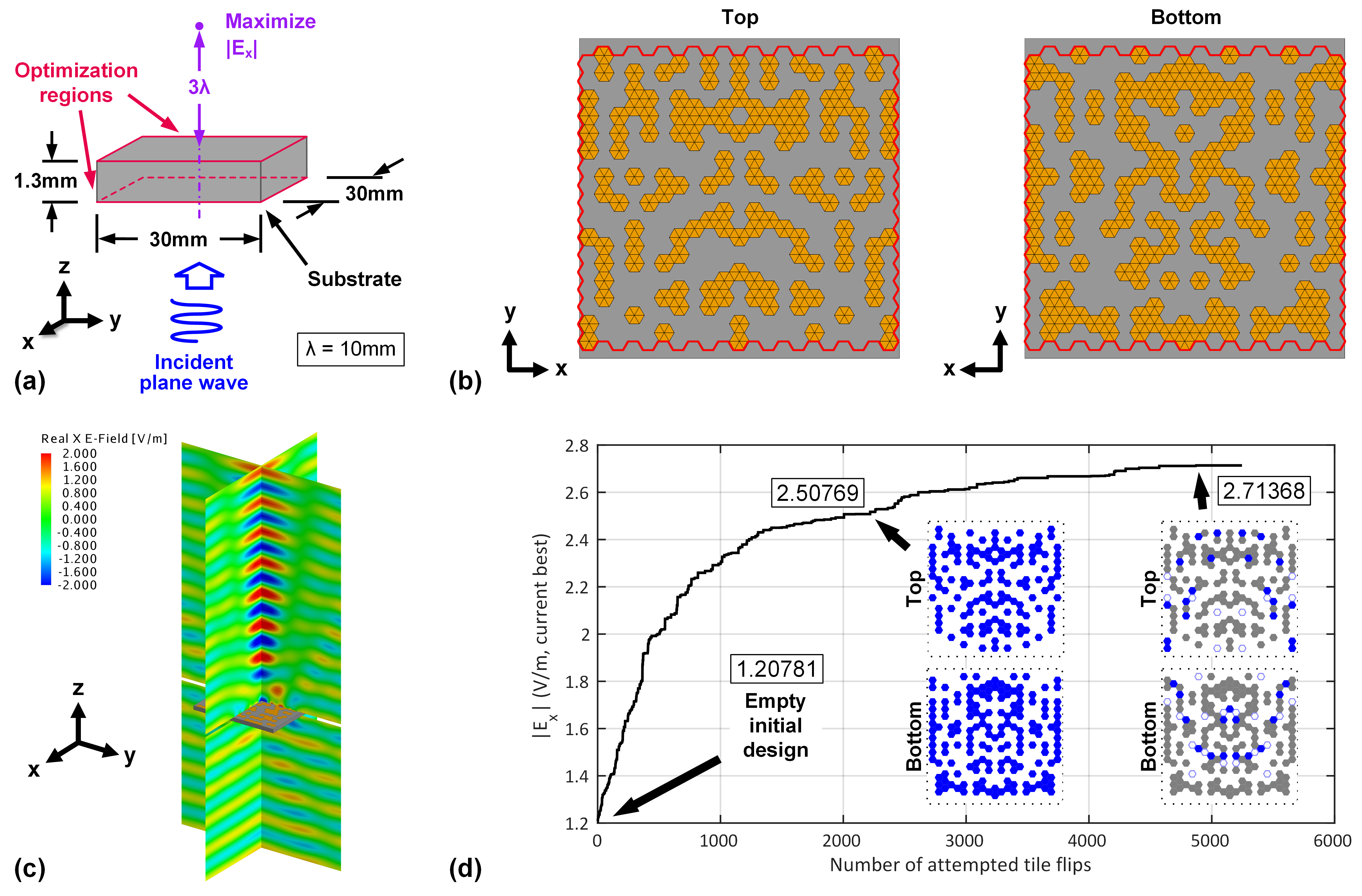}
\caption{
\textbf{30GHz two-layer beamforming metasurface.} (a) Illustration of simulation environment and optimization objective; (b) Final optimized device design; (c) Visualization of the electric field distribution resulting from a 30GHz plane wave incident on the metasurface; (d) Evolution of objective function during inverse design.
}
\addcontentsline{toc}{subsubsection}{Supplementary Figure 3}
\label{fig:supp_bfmsurf}
\end{figure}

The beamforming metasurface design is optimized at 30GHz. The substrate ($\varepsilon_r=3.66$) represents a Rogers RO4350 high-frequency laminate, with a width and height of $30$mm each (corresponding to $3\lambda \times 3\lambda$ in free space) and a thickness of $1.3$mm (corresponding to a quarter wavelength inside the dielectric). The incident excitation is an x-polarized plane wave traveling in the +z direction, and the objective function seeks to maximize $|E_x|$ at a point $30$mm ($3\lambda$) above the top (+z) of the metasurface.
Due to the flexibility of MoM in supporting non-Cartesian triangular meshes, arbitrarily shaped tiles can be used. To demonstrate this, we use hexagonal tiles in this example, which have the added advantage that any two adjacent tiles must share at least one full edge and cannot be touching only at a single corner point (which is possible when using square tiles). Two $21 \times 17$ grids of hexagonal tiles are used, one on each side of the substrate, resulting in a total of $714$ optimization parameters. Each tile has an edge length of 0.9375mm. As with the examples in the main text, the Direct Binary Search (DBS) optimization algorithm is used, and each optimization parameter (element in the design matrix $P$) can assume values of $0$ (empty) or $1$ (PEC).

The metasurface design, the electric field distribution simulated with FEKO, and the evolution of the objective function as optimization progresses are shown in Supplementary Fig. \ref{fig:supp_bfmsurf}. Note that the hexagonal tile map has a constant number of tiles in each column, resulting in staggered columns and only a single axis of symmetry. Precomputation is performed with FEKO utilizing the default MoM-SEP solver with single precision on an AMD EPYC 7763 node with 128 cores, taking 82.9s. On the same hardware, PNGF optimization takes approximately 45.5s in total, including constructing the $C$ matrix, evaluating the initial objective function, and 5,246 total attempted tile flips. 
This corresponds to a speedup of 8,370x compared to using FEKO to rerun the full-wave simulation at each iteration of the optimization, which would require approximately 106 hours (5,246 $\times$ 72.6s for solving the final design with FEKO).

\newpage
\subsection{Numerical accuracy}
\label{supp:numerical_accuracy}

The following tables present the numerical values obtained for the magnitude of $S_{11}$, as well as the directivity $D(\theta, \phi)$ at $(\theta, \phi) = (0,0)$ at each optimization frequency for the substrate antenna design example. Values obtained from PNGF (i.e., the $\mathbf{x_{obj}}$ values for the final design, found at the last optimization of iteration that yields a performance improvement) are compared against results from a simulation of the final substrate antenna design with the finite-difference frequency domain (FDFD) solver used for precomputation as well as the finite-difference time domain (FDTD) results from Fig. \ref{fig:subant}(b). From the FDTD simulation, the quantities are obtained using a discrete-time Fourier transform (DTFT) to find the necessary field values in the frequency domain at each frequency of optimization.

The values obtained with PNGF agree with those from the FDFD solver to approximately 4 digits of accuracy when single precision is used for MUMPS in the FDFD solver (for simulating the final design as well as for precomputing the numerical Green's function matrices used for PNGF). When double precision is used, approximately 10 digits of agreement are obtained. 
The precomputation timing results reported in Table \ref{tb:performance} for APF utilize single precision. When double precision is used with the same computing resources, the precomputation time for this design example increases to 14.7min. 
It should be noted that the same discretization is utilized for both FDFD and FDTD, and differences between the results arise solely due to numerical errors introduced by discretizing time and utilizing a finite time step with the time-domain solver.

\begin{table}[h]
\caption{$S_{11}$ magnitudes of the substrate antenna design obtained with PNGF, FDFD, and FDTD.}
\label{tb:supp_numacc_s11}
\begin{tabular}{*6c}
    \toprule
    \multirow{2}{*}{$\begin{matrix}\textbf{Frequency} \\ \textbf{(GHz)}\end{matrix}$}
     & \multirow{2}{*}{\textbf{Solver}} 
        & \multicolumn{2}{c}{\textbf{Single precision}} & \multicolumn{2}{c}{\textbf{Double precision}} \\
        & &  ${|S_{11}|}$ & {Error}$^{\;1}$ & ${|S_{11}|}$ & {Error}$^{\;1}$ \\
    \toprule 
    \multirow{3}{*}{25} 
        & PNGF 
            & 0.2139935 & - 
            & 0.2139986427183314 & - \\
        & FDFD 
            & 0.2139607 & 1.53$\times 10^{-4}$ 
            & 0.2139986427061767 & 5.68$\times 10^{-11}$ \\
        & FDTD$^{\;2}$
            & - & -
            & 0.2139941816777345 & $2.08\times 10^{-5}$ \\
    \midrule
    \multirow{3}{*}{27.5} 
        & PNGF 
            & 0.1156671 & - 
            & 0.1156679743901394 & - \\
        & FDFD 
            & 0.1156317 & 3.06$\times 10^{-4}$ 
            & 0.1156679744284358 & 3.31$\times 10^{-10}$ \\
        & FDTD$^{\;2}$
            & - & -
            & 0.1157132553502266 & 3.91$\times 10^{-4}$ \\
    \midrule
    \multirow{3}{*}{30} 
        & PNGF 
            & 0.09093406 & - 
            & 0.09093572688067568 & - \\
        & FDFD 
            & 0.09096325 & 3.21$\times 10^{-4}$ 
            & 0.09093572688961719 & 9.83$\times 10^{-11}$ \\
        & FDTD$^{\;2}$
            & - & -
            & 0.09092127626013229 & 1.59$\times 10^{-4}$ \\
    \midrule
    \multirow{3}{*}{32.5} 
        & PNGF 
            & 0.1222233 & -
            & 0.1222178408501603 & - \\
        & FDFD 
            & 0.1220848 & 1.13$\times 10^{-3}$
            & 0.1222178408533571 & 2.62$\times 10^{-11}$ \\
        & FDTD$^{\;2}$
            & - & -
            & 0.1222471707081718 & 2.40$\times 10^{-4}$ \\
    \midrule
    \multirow{3}{*}{35} 
        & PNGF 
            & 0.1346681 & - 
            & 0.1346679204155417 & - \\
        & FDFD 
            & 0.1347649 & 7.18$\times 10^{-4}$
            & 0.1346679204363667 & 1.55$\times 10^{-10}$ \\
        & FDTD$^{\;2}$
            & - & -
            & 0.1346542107180774 & 1.02$\times 10^{-4}$ \\
    \botrule
\end{tabular}
\footnotetext[1]{Relative error of the PNGF result with the solver in each row as the reference.}
\footnotetext[2]{FDTD simulation was performed with double precision only.}
\end{table}

\begin{table}[t]
\caption{Directivity of the substrate antenna design obtained with PNGF, FDFD, and FDTD.}
\label{tb:supp_numacc_dir}
\begin{tabular}{*6c}
    \toprule
    \multirow{2}{*}{$\begin{matrix}\textbf{Frequency} \\ \textbf{(GHz)}\end{matrix}$}
     & \multirow{2}{*}{\textbf{Solver}} 
        & \multicolumn{2}{c}{\textbf{Single precision}} & \multicolumn{2}{c}{\textbf{Double precision}} \\
        & &  $D(0,0)^{\;1}$ & {Error}$^{\;2}$ & $D(0,0)^{\;1}$ & {Error}$^{\;2}$ \\
    \toprule 
    \multirow{3}{*}{25} 
        & PNGF 
            & 7.811967 & - 
            & 7.812037601247257 & - \\
        & FDFD 
            & 7.812057 & 1.15$\times 10^{-5}$ 
            & 7.812037601343792 & 1.24$\times 10^{-11}$ \\
        & FDTD$^{\;3}$
            & - & -
            & 7.815307868960594 & $4.18\times 10^{-4}$ \\
    \midrule
    \multirow{3}{*}{27.5} 
        & PNGF 
            & 10.83782 & - 
            & 10.83784770721897 & - \\
        & FDFD 
            & 10.83833 & 4.71$\times 10^{-5}$ 
            & 10.83784770700359 & 1.99$\times 10^{-11}$ \\
        & FDTD$^{\;3}$
            & - & -
            & 10.83938138413976 & 1.41$\times 10^{-4}$ \\
    \midrule
    \multirow{3}{*}{30} 
        & PNGF 
            & 11.99033 & - 
            & 11.99031145686420 & - \\
        & FDFD 
            & 11.98917 & 9.68$\times 10^{-5}$ 
            & 11.99031145721867 & 2.96$\times 10^{-11}$ \\
        & FDTD$^{\;3}$
            & - & -
            & 11.99115781408769 & 7.06$\times 10^{-5}$ \\
    \midrule
    \multirow{3}{*}{32.5} 
        & PNGF 
            & 14.33541 & -
            & 14.33530754234066 & - \\
        & FDFD 
            & 14.33493 & 3.35$\times 10^{-5}$
            & 14.33530754248159 & 9.83$\times 10^{-12}$ \\
        & FDTD$^{\;3}$
            & - & -
            & 14.33433344364580 & 6.80$\times 10^{-5}$ \\
    \midrule
    \multirow{3}{*}{35} 
        & PNGF 
            & 14.71850 & - 
            & 14.71859356714017 & - \\
        & FDFD 
            & 14.71970 & 8.15$\times 10^{-5}$
            & 14.71859356731010 & 1.15$\times 10^{-11}$ \\
        & FDTD$^{\;3}$
            & - & -
            & 14.71644964221402 & 1.46$\times 10^{-4}$ \\
    \botrule
\end{tabular}
\footnotetext[1]{Values are given in absolute (linear) scale.}
\footnotetext[2]{Relative error of the PNGF result with the solver in each row as the reference.}
\footnotetext[3]{FDTD simulation was performed with double precision only.}
\end{table}

\clearpage
\newpage

\newpage
\subsection{Objective functions for design studies}
\label{supp:objfuncs}

Example objective functions for each of the design studies are presented in the following sections:

\subsubsection{Substrate antenna}

For the substrate antenna, it is desirable to maximize the directivity in the direction broadside to the antenna over a wide frequency range while maintaining good impedance-matching performance. To accomplish this,
\begin{equation}
\begin{split}
f_{\text{obj}, i} = 
a_1 \left|S_{11,i}\right|^2 + \left(a_2 - D_i(0, 0) \right)^2
\end{split}
\label{eq:supp_subant_objfunc}
\end{equation}
may be minimized over a set of $N$ frequencies. The frequencies of optimization are indexed with $i$, $S_{11,i}$ is the reflection coefficient at the $i$th frequency, and $D_{i}(\theta, \phi)$ is the directivity. The parameters $a_1$ and $a_2$ are empirical constants which may be adjusted for the best optimization results. The objectives at all the frequencies may be combined into a single scalar objective function using a suitable weighting method, such as using a harmonic or arithmetic mean or a min-max approach.
For the substrate antenna design shown in Fig. \ref{fig:subant}, optimization is performed with five frequencies \{25, 27.5, 30, 32.5, 35GHz\}.

\subsubsection{Switched-beam antenna}

For this design study, the directivities in the target directions with the switch open and closed should be maximized while maintaining good impedance-matching performance. An example objective function to achieve this is
\begin{equation}
\begin{split}
f_{\text{obj}, i} =
& \: a_1\left[\left|S_{11,i}^\text{on}\right|+\left|S_{11,i}^\text{off}\right| \right] 
+ a_2 \left[\frac{1}{D_{i}^\text{on}(45^\circ,270^\circ)}+\frac{1}{D_{i}^\text{off}(45^\circ,90^\circ)} \right] \\
&+a_3 \left[D_{i}^\text{on}(45^\circ,90^\circ)+D_{i}^\text{off}(45^\circ,270^\circ) \right]
,
\end{split}
\label{eq:supp_sba_objfunc}
\end{equation}
where $S_{11,i}^\text{on}$ and $S_{11,i}^\text{off}$ are the reflection coefficients with the switch closed and open, respectively, at the $i$th frequency, and $D_{i}^\text{on}(\theta, \phi)$ and $D_{i}^\text{off}(\theta, \phi)$ are the directivities in each case. 
The target directions correspond to angles $\pm 45^\circ$ from a vector normal to the surface of the antenna (i.e. z-axis in Fig. \ref{fig:sba}), and it is desirable to maximize $D_{i}^\text{on}$ and minimize $D_{i}^\text{off}$ for $(\theta = 45^\circ, \phi = 270^\circ)$, and vice versa for $(\theta = 45^\circ, \phi = 90^\circ)$.
As before, the optimizer seeks to minimize $f_{\text{obj}, i}$, and the parameters $a_1$, $a_2$, and $a_3$ may be found empirically. Optimization for the switched-beam antenna design shown in Fig. \ref{fig:sba} is carried out with three frequencies \{29.5, 30, 30.5GHz\}.

\subsubsection{Substrate-integrated waveguide}

For the substrate-integrated waveguide structure, it is desirable to minimize the insertion loss, which may be done by minimizing
\begin{equation}
\begin{split}
f_{\text{obj}, i} = (1 - |S_{21, i}|)^2,
\end{split}
\label{eq:supp_siw_objfunc}
\end{equation}
where $S_{21, i}$ is the transmission coefficient at frequency $i$. The design shown in Fig. \ref{fig:siw} is optimized at five frequencies \{11, 11.5, 12, 12.5, 13GHz\}.

\newpage
\subsection{Design study simulation results comparison}
\label{supp:designssims}

Enlarged versions of the performance result plots shown in Fig. \ref{fig:sba} and Fig. \ref{fig:siw} in the main text are presented below with additional simulation results from CST. The feed connector is not modeled in the CST results. HFSS results without the connector are also included for the SIW design.

\begin{figure}[!ht]
\centering
\includegraphics[width=\textwidth,trim=10 10 10 10,clip]{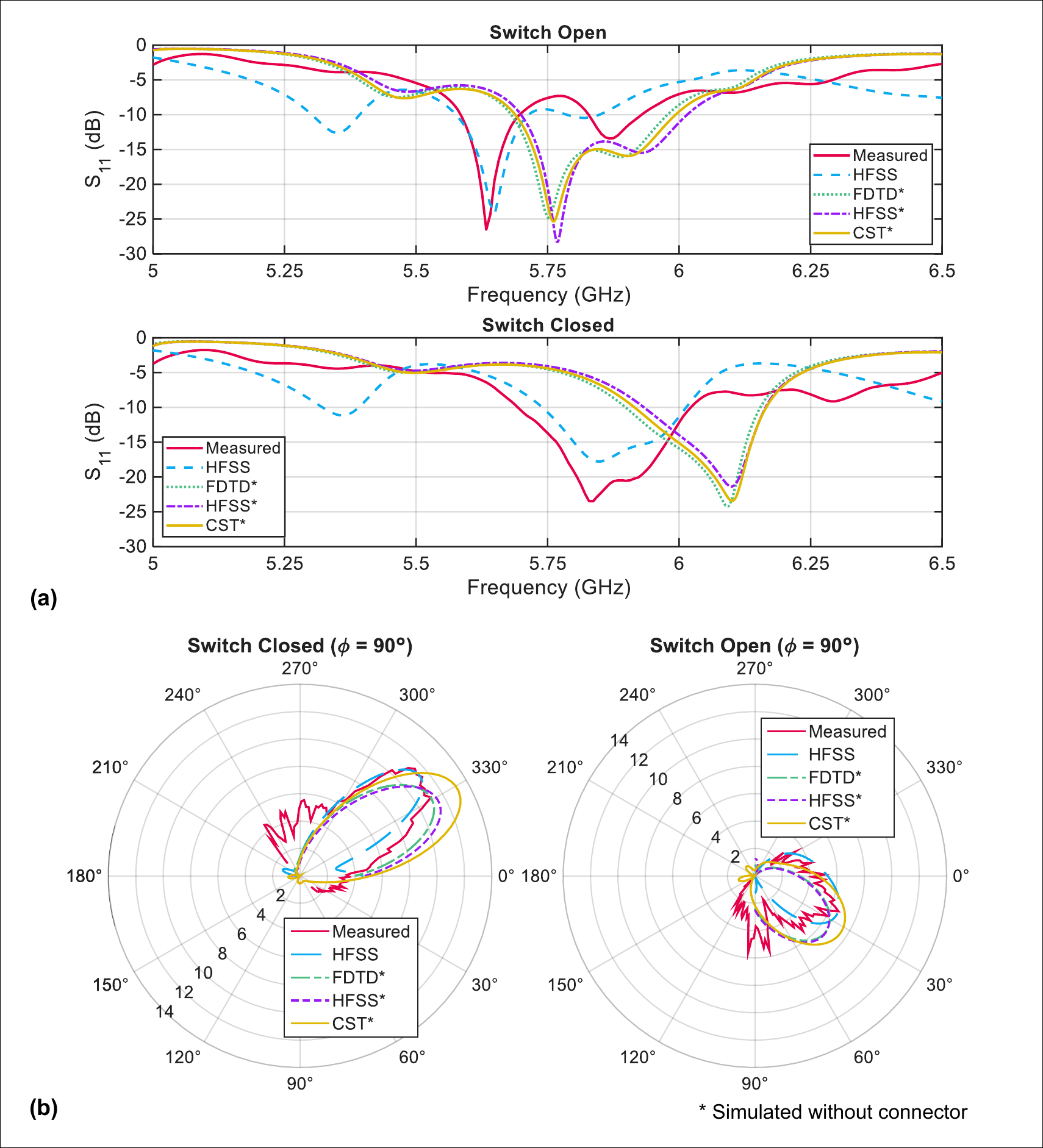}
\caption{\textbf{Switched-beam antenna design performance.} (a) Simulated and measured $S_{11}$; (b) Simulated and measured radiation patterns at 6GHz in linear scale relative to an ideal isotropic radiator.}
\addcontentsline{toc}{subsubsection}{Supplementary Figure 4}
\label{fig:supp_sba}
\end{figure}

\begin{figure}[!ht]
\centering
\includegraphics[width=\textwidth]{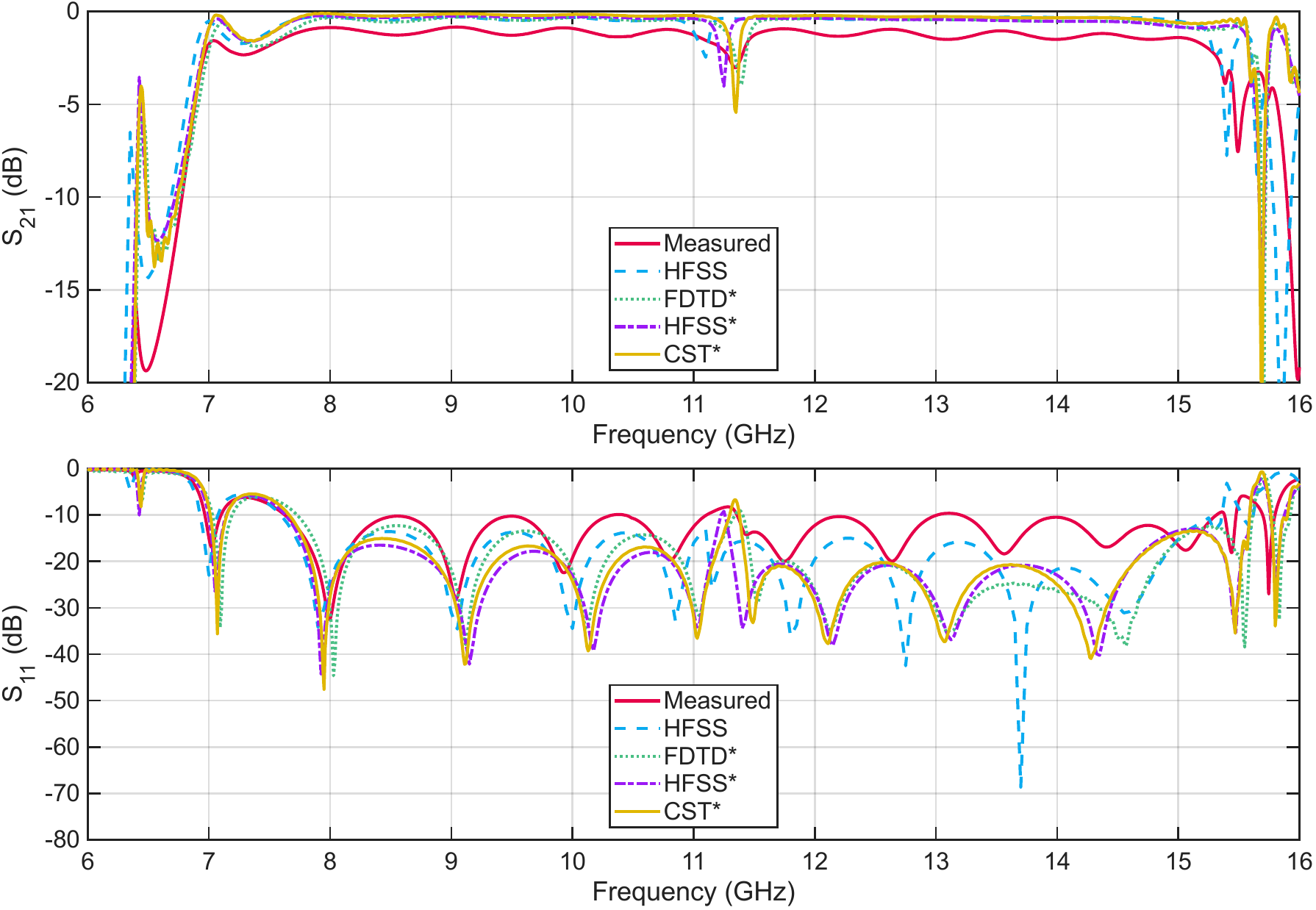}
\caption{\textbf{Substrate-integrated waveguide design performance.} Simulated and measured $S_{11}$ and $S_{21}$.}
\addcontentsline{toc}{subsubsection}{Supplementary Figure 5}
\label{fig:supp_siw}
\end{figure}

\newpage
\vspace{1cm}
\newpage
\color{black}


\end{document}